\newif\ifpdf 
\ifx\pdfoutput\undefined 
  \pdffalse 
\else 
  \pdftrue 
\fi

\newif\ifpreprint
\preprinttrue

\ifpreprint
  \documentclass[numberedappendix]{emulateapj}
  \usepackage[varg]{txfonts}
\else
  \documentclass[12pt,preprint]{aastex}
\fi

\usepackage[latin1]{inputenc}

\ifpreprint

  \let\url\relax
  \ifpdf
    \usepackage[pdftex]{hyperref}
  \else
    \usepackage{hyperref}
  \fi
  \usepackage[all]{hypcap}
  \makeatletter
  \renewcommand{\fps@figure}{tp}
  \makeatother
\else
  
  \makeatletter
  \renewcommand{\fps@figure}{p}
  \makeatother
\fi

\ifpreprint
  \setkeys{Gin}{width=\linewidth,height=0.9\textheight,keepaspectratio=true}
\else
  \setkeys{Gin}{width=\linewidth,height=0.7\textheight,keepaspectratio=true}
\fi

\ifpreprint
  \newcounter{ionstage}
  \renewcommand{\ion}[2]{\setcounter{ionstage}{#2}%
    \ensuremath{\mathrm{#1\,\scriptstyle\Roman{ionstage}}}}
\fi

\newcommand\hii{\ion{H}{2}}
\newcommand\ifront{ionization front}
\newcommand\Sub[1]{_\mathrm{#1}}
\newcommand\Unit[1]{\mathrm{#1}}
\newcommand\cm{\Unit{cm}}
\newcommand\parsec{\Unit{pc}}
\newcommand\pcc{\Unit{cm^{-3}}}
\newcommand\persecond{\Unit{s^{-1}}}

\newcommand\Kelvin{\Unit{K}}
\newcommand\years{\Unit{yr}}
\newcommand\kms{\Unit{km\,s^{-1}}}
\newcommand\punits{\Unit{dyne\,cm^{-2}}}
\newcommand\Angstrom{\Unit{\AA}}
\newcommand\OIlam{[\ion{O}{1}]\,6300\,\AA\@}
\newcommand\NIIlam{[\ion{N}{2}]\,6584\,\AA\@}
\newcommand\OIIIlam{[\ion{O}{3}]\,5007\,\AA\@}
\newcommand\dfrac[2]{\frac{\displaystyle#1}{\displaystyle#2}}
\newcommand\thC{$\theta^1$~Ori~C\@}

\ifpreprint
  \graphicspath{
    {.},
  }
\else
  \graphicspath{
    {/home/wjh/Orion2004/Papers/A1/FlatFigs/},
  }
\fi

\begin{document}

\title{Photoevaporation Flows in Blister \hii{} Regions:\\
  I.~Smooth Ionization Fronts and Application to the Orion Nebula}

\shorttitle{Flows from Smooth Ionization Fronts}

\author{W. J. Henney\altaffilmark{1}, S. J. Arthur\altaffilmark{1},
  and Ma.~T. García-Díaz}

\affil{Centro de Radioastronomía y Astrofísica, UNAM Campus Morelia,
  Apdo.\@ Postal 3-72, 58090 Morelia, Michoacán, México.}

\email{w.henney,j.arthur,t.garcia@astrosmo.unam.mx} 

\altaffiltext{1}{Work carried out in part while on sabbatical at the
  Department of Physics and Astronomy, University of Leeds, LS2~9JT,
  UK.}

\shortauthors{Henney, Arthur, \& García-Díaz}

\begin{abstract}
  We present hydrodynamical simulations of the photoevaporation of a
  cloud with large-scale density gradients, giving rise to an ionized,
  photoevaporation flow. The flow is found to be approximately steady
  during the large part of its evolution, during which it can resemble
  a ``champagne flow'' or a ``globule flow'' depending on the
  curvature of the ionization front. The distance from source to
  ionization front and the front curvature uniquely determine the
  structure of the flow, with the curvature depending on the steepness
  of the lateral density gradient in the neutral cloud.

  We compare these simulations with both new and existing observations
  of the Orion nebula and find that a model with a mildly convex
  ionization front can reproduce the profiles of emission measure,
  electron density, and mean line velocity for a variety of emitting
  ions on scales of $10^{17}$--$10^{18}\,\cm$. The principal failure
  of our model is that we cannot explain the large observed widths of
  the [\ion{O}{1}]~6300\,\AA{} line that forms at the ionization
  front.  
\end{abstract}


\section{Introduction}
\label{sec:introduction}

\hii{} regions, formed by the action of ionizing photons from a
high-mass star on the surrounding gas, often occur in highly
inhomogeneous environments. Most \hii{} regions are found close to
high density concentrations of molecular gas and, for those regions
that are optically visible, there is a tendency for the ionized
emission to be blueshifted with respect to the molecular emission
\citep{1978A&A....70..769I}. Since optically visible \hii{} regions
must be on the near side of the molecular gas, this implies that the
ionized gas is flowing away from the molecular cloud. The archetypal
and best-studied example of such a ``blister-type'' \hii{} region is
the nearby Orion nebula, which is ionized by the O7~V star, \thC{}
\citep[for recent reviews, see][]{2001PASP..113...29O,
  2001PASP..113...41F, 2001ARA&A..39...99O}.  This compact \hii{}
region, which lies in front of the molecular cloud OMC~1, shows a
bright core of ionized emission, approximately $10^{18}\,\cm$ in
diameter, surrounded by a fainter halo that extends out to distances
of order $10^{19}\,\cm$ \citep{2001AJ....121..399S}.  Surprisingly,
the thickness along the line of sight of the layer responsible for the
bulk of the ionized emission in the core of the nebula is only $\sim 2
\times 10^{17}\,\cm$ \citep[e.g.,][]{1991ApJ...374..580B,
  1995ApJ...438..784W}, which is significantly smaller than both the
lateral extent of the core and the deduced distance between the
ionizing star and the front, which is also of order $10^{18}\,\cm$.
Three-dimensional reconstruction of the shape of the nebula's
principal ionization front \citep{1995ApJ...438..784W} shows the front
to have a saddle-shaped structure, with a slight concavity in the NS
direction and a slight convexity in the EW direction \citep[see][
Figure~3]{2001ARA&A..39...99O}. In both cases the radius of curvature
is large compared with star-front distance, although there are also
many smaller scale irregularities in the front, with the most
prominent of these being associated with the ``Bright Bar'' in the SE
region of the nebula. Extensive spectroscopic studies of the Orion
nebula have been carried out to determine the velocity structure in
the ionized gas \citep{1967ApJ...148..925K, 1974PASP...86..616B,
  1979A&A....75...34P, 1993ApJ...403..678O, 1999AJ....118.2350H,
  2001ApJ...556..203O, 2004Doi-kinematics}.  Although these show a
wealth of fine-scale structure, the fundamental result is that lines from
the more highly ionized species such as H$\alpha$ and [\ion{O}{3}] are
blue-shifted by approximately $10\,\kms$ with respect to the molecular
gas of OMC-1, which lies behind the nebula.  Lines from
intermediate-ionization species such as [\ion{S}{2}] and [\ion{N}{2}]
being found at intermediate velocities. All these considerations lead
to a basic model of the nebula as a thin layer of ionized gas that is
streaming away from the molecular cloud, as was first proposed by
\citet{1973ApJ...183..863Z}.

Despite, or perhaps because of, the mass of observational data that
has been accumulated on the Orion nebula, no attempt has been made to
develop a self-consistent dynamical model of the ionized flow since
the early work of \citet{1963AJ.....68R.296V, 1964ApJ...139..869V}.
Most modern models of the nebula have concentrated great effort on
details of the atomic physics and radiative transfer
\citep{1991ApJ...374..580B, 1991ApJ...374..564R} and, while these have
been quite successful in explaining the ionization structure of the
nebula, they implicitly assume a static configuration and are forced
to adopt ad hoc density configurations that have no physical basis.
The importance of photoevaporation flows has been recognized in the
context of the modeling of other \hii{} regions
\cite{1996AJ....111.2349H, 1998AJ....116..163S, 2000ApJ...535..847S,
  2002AJ....124.3305M} but again no attempt has been made to treat the
dynamics in a consistent manner.  More recently, steady-state dynamic
models of weak-D ionization fronts have been developed and applied to
the Orion nebula \citep{2005ApJ...621..328H}, with some success in
explaining the structure of the Bright Bar, where the ionization front
is believed to be viewed edge-on. However, the one-dimensional,
plane-parallel nature of these models renders them incapable of
capturing the global geometry of the nebula and indeed they completely
fail to reproduce the observed correlation between line velocity and
ionization potential.

The evolution of an \hii{} region in an inhomogeneous medium was first
studied by \citet{1979A&A....71...59T}, who considered the
``break-out'' of a photoionized region from the edge of a dense cloud,
such as would occur if a high-mass star formed inside, but close to
the edge of, a molecular cloud.  Once the ionization front reaches the
edge of the cloud, it rapidly propagates through the low-density
intercloud medium, followed by a strong shock driven by the higher
pressure ionized cloud material. At the same time, a rarefaction wave
propagates back into the ionized cloud, giving rise to a ``champagne
flow'' in which the \hii{} region is ionization-bounded on the high
density side and density-bounded on the low-density side and the
ionized gas flows down the density gradient. This work has been
extended in many subsequent papers using both one-dimensional and
two-dimensional numerical simulations \citep{1979ApJ...233...85B,
  1981A&A....98...85B, 1982A&A...108...25Y, 1989RMxAA..18...65F,
  1997A&A...326.1195C}, which generally concentrate on non-steady
aspects of the flow. 

Another scenario that has been extensively studied is the ionization
of a neutral globule by an external radiation field
\citep{1956BAN....13...77P, 1968Ap&SS...1..388D, 1981A&A....99..305T,
  1984A&A...135...81B, 1989ApJ...346..735B,1990ApJ...354..529B}, to
form what is variously referred to as an ionized photoevaporation or
photoablation flow. In this case, after an initial transient phase in
which the globule is compressed, the ionization front propagates very
slowly into the globule and the ionized flow is steady to a good
approximation.  Although originally developed in the context of bright
rims and globules in \hii{} regions, similar models have more recently
been applied to the photoevaporation of circumstellar accretion disks
\citep{1998ApJ...499..758J, 1998AJ....116..322H} and cometary knots in
planetary nebulae \citep{2001ApJ...548..288L, 2000AJ....119.2910O}. In
all these cases, the ionization front has positive (convex) curvature,
and the ionized gas accelerates away from the ionization front in a
transonic, spherically divergent flow.

In a strictly plane-parallel geometry, a time-stationary champagne
flow cannot exist. This is because, in the absence of body forces, the
acceleration of a steady plane-parallel flow is proportional to the
gradient of the sound speed, which is roughly constant in the
photoionized gas.  However, it has recently been shown
\citep{2003RMxAC..15..175H} that a steady, ionized, isothermal,
transonic, divergent flow can exist even if the ionization front is
flat, so long as the ionizing source is at a finite distance from the
front. In such a case, the divergence of the radiation field breaks
the plane-parallel geometry and induces a (weaker) divergence in the
ionized flow, allowing it to accelerate (an expanded and corrected
version of this derivation is presented in
Appendix~\ref{sec:analytic-model-flow} below). It is natural to
speculate that a steady flow may also be possible even if the front
has negative (concave) curvature, so long as the radius of curvature
of the front is greater than its distance from the ionizing source.
By this means, the champagne flow problem can be effectively reduced
to a modified version of the globule flow problem.

In order to investigate this possibility, we here present
hydrodynamical simulations of the photoionization of a neutral cloud
with large-scale density gradients and develop a taxonomy of the
resulting \hii{} region flow structure and dynamics. We concentrate on
the quasi-stationary stage of the \hii{} region evolution, in which
the structure changes slowly compared with the dynamical timescale of
the ionized flow, with the result that the flow is approximately
steady in the frame of the front. Depending on the curvature of the
ionization front, our models can resemble both ``blister flows''
\citep{1978A&A....70..769I} if the curvature is negative (concave), or
``bright rims'' \citep{1956BAN....13...77P}, if the curvature is
positive (convex). The algorithm, initial conditions, and parameters
of these numerical simulations are described in
\S~\ref{sec:numer-simul} and their results are set out in
\S~\ref{sec:model-results}. In
\S~\ref{sec:observational-data} we derive empirical parameters
for the Orion nebula that will allow us to make meaningful comparisons
with our models. These comparisons are carried out in
\S~\ref{sec:comp-with-simul}. In \S~\ref{sec:discussion} we
discuss the implications of our model comparisons to Orion and outline
some shortcomings of the present simulations that will be addressed in
future work. In \S~\ref{sec:conclusions} we draw the conclusions
of our paper. Finally, we present some more technical material in two
appendices. The first derives an approximate analytic model for the
flow from a plane ionization front, while the second investigates the
question of whether the ionization fronts in our simulations are
\hbox{D-critical}.

\section{Numerical Simulations}
\label{sec:numer-simul}

In this section we present time-dependent numerical hydrodynamic
simulations of the photoionization of neutral gas by a point source
offset from a plane density interface.

\subsection{Implementation Details}
\label{sec:details}

The Euler equations are solved in two-dimensional cylindrical
geometry, using a second-order, finite-volume, hybrid scheme, in which
alternate timesteps are calculated by the Godunov method and the
Van~Leer flux-vector splitting method
\citep{Godunov,vanLeer1982}.\footnote{The velocity shear in the
  Van~Leer steps helps avoid the ``carbuncle instability'', which
  occurs in pure Godunov schemes when slow shocks propagate parallel
  to the grid lines \citep{Quirk1994,Sanders1998}.}  The radiative
transfer of the ionizing radiation is carried out by the method of
short characteristics \citep{1999RMxAA..35..123R}, with the only
source of ionizing opacity being photoelectric absorption by neutral
hydrogen.  Only one frequency of radiation is included but
radiation-hardening is accounted for by varying the effective
photoabsorption cross-section as a function of optical depth
\citep{2005ApJ...621..328H}. The diffuse field is treated by the on-the-spot
approximation. Dust opacity is not explicitly included in our
calculations but is treated to zeroth order by merely reducing the
stellar ionizing luminosity to account for absorption by grains. As
shown in Appendix~\ref{sec:analytic-model-flow}, this is a rather good
approximation at least in the case of a flow from a plane ionization
front. 

Ionization and recombination of hydrogen is calculated by exact
integration of the relevant equations over a numerical timestep.
Heating and cooling of the gas are calculated by a second-order
explicit scheme. The only heating process considered is the
photoionization of hydrogen, with the energy deposited per ionization
increasing with optical depth according to the hardening of the
radiation field. The most important cooling processes included are
collisional excitation of Lyman alpha, collisional excitation of
neutral and ionized metal lines (assuming standard ISM abundances),
and hydrogen recombination. Other cooling processes included in the
code, such as collisional ionization and bremsstrahlung, are only
important at higher temperatures than are found in the current
simulations.

\subsection{Initial Conditions}
\label{sec:initial-conditions}

At the beginning of the simulation, the ionizing source is placed at
$r = 0$, $z = z_0$, and the computational domain is filled with
neutral gas with a separable density distribution:
\begin{equation}
  \label{eq:rho-sep}
  \rho(r,z) = \rho_0  F(r) G(z) . 
\end{equation} 
For comparison with the analytic model of
Appendix~\ref{sec:analytic-model-flow}, we wish to contrive a neutral
density distribution that will ensure that the ionization front stays
approximately flat throughout its evolution, although we will also
investigate the flow from curved fronts.

The initial flatness of the ionization front is guaranteed by choosing
$G(z)$ to be a step function between a very low value (that will be
ionized in a recombination time) and a very high value (that will trap
the ionization front at the interface). In practice, we choose the
function
\begin{equation}
  \label{eq:rho-z}
  \log_{10} G(z) = \delta \tanh((z-z_1)/h_1) , 
\end{equation}
where $h_1$ is the thickness of the interface, which occurs at a
height $z_1$ ($> z_0$). The density contrast between the two sides is
$10^{2\delta}$. 

In order for the ionization front to remain flat as it propagates into
the dense neutral gas above the interface, we let the initial density
fall off gradually in the radial direction according to the function
\begin{equation}
  \label{eq:2}
  F(r) = \left[1 + \left(\frac{r}{z_1-z_0}\right)^2\right]^{-\alpha} , 
\end{equation}
where $\alpha$ is an adjustable parameter controlling the steepness of
the distribution. The analytic model of
Appendix~\ref{sec:analytic-model-flow} suggests that $\alpha = 1$
should ensure that all parts of the ionization front propagate at the
same speed. Choosing smaller values of $\alpha$ should mean that the
front propagates fastest along the symmetry axis and so the front
becomes concave with time. Conversely, larger values of $\alpha$ will
cause the front to propagate faster away from the axis and so become
convex with time.

A characteristic temperature, $T_0$, is chosen, which corresponds to
gas at a density of $\rho_0$. The temperature distribution throughout
the grid is then set so as to give a constant pressure where $\rho >
\rho_0$ and a constant temperature where $\rho < \rho_0$.\footnote{We
  require a constant pressure in the high-density, undisturbed gas so
  that it remains motionless throughout the simulation. The low
  density gas is designed to be immediately ionized and then swiftly
  swept from the grid by the photoevaporation flow. A constant
  temperature ensures that this happens as smoothly as possible.}

In this paper we follow \citet{1992phas.book.....S} in denoting gas
velocities in the frame of reference of the ionizing star by $u$, gas
velocities in the frame of reference of the ionization front by $v$,
and pattern speeds of ionization, heating, and shock fronts by $U$.

\subsection{Model Parameters}
\label{sec:model-parameters}

We present three different models, which illustrate the effect of
different radial density distributions in the neutral gas, with
$\alpha = 0$ (Model~A), $\alpha = 1$ (Model~B), and $\alpha = 2$
(Model~C). All other parameters are identical between the three models
and are chosen to be representative of a compact \hii{} region such as
the Orion nebula. The simulation is carried out on a $512\times512$
grid, with sides of length $1\,\parsec$. The source has an ionizing
luminosity $Q\Sub{H} = 10^{49} \,\persecond $, with a
$40,000\,\Kelvin$ black-body spectrum, and is placed half-way up the
simulation grid on the $z$-axis.  The characteristic density for the
simulation is set at $\rho_0 = 10^4 m \,\pcc$, where $m$ is the mean
mass per nucleon (assumed to be $1.3 m\Sub{H}$). The separation,
$z_1-z_0$, between the source and the density interface is chosen so
that the analytic steady-state model of
Appendix~\ref{sec:analytic-model-flow} below would give a peak ionized
density of $\rho_0/m = 10^{4}\,\pcc$ in the photoevaporation flow.
This yields $z_1-z_0 = 4.8\times 10^{17}\,\cm = 0.155\,\parsec$. The
initial density contrast across the interface is set at $\delta = 2$,
yielding $n = 10^6 \,\pcc$ on the high-density side and $n = 10^2
\,\pcc$ on the low-density side. The thickness of the interface is set
at $h_1 = 0.3(z_1-z_0)$ and the characteristic temperature  is
set at $T_0 = 600\,\Kelvin$.\footnote{Note that this implies
  temperatures as low as $10\,\Kelvin$ in the neutral gas. We do not
  attempt to model the heating/cooling or molecular processes in the
  cold gas.}

In order to make the simulations feasible, it is necessary to use a
hydrogen photoionization cross-section that is considerably smaller
than the physically correct value, which would be roughly $\sigma_0 =
3\times10^{-18}\,\cm^2$ for unhardened radiation from a
$40,000\,\Kelvin$ black body \citep[Appendix~A]{2005ApJ...621..328H}.  In
these models, we instead use a value of $\sigma_0 = 5\times
10^{-20}\,\cm^2$, which is 60 times smaller.  This is necessary in
order to properly resolve the deep part of the ionization front (at
$\tau \simeq 10$) where the initial heating of the photoevaporation
flow occurs. Failure to resolve this region results in spurious
large-magnitude oscillations in the flow. Using a smaller value of
$\sigma_0$ increases the width of the ionization front but has a
negligible effect on the kinematic behavior of the gas once it is
fully ionized.  We have studied the effect of increasing or decreasing
$\sigma_0$ and find that our results are generally not affected, with
the exception of some specific points noted below.

Although each model is calculated for specific values of the length
scale and ionizing luminosity, they may be safely rescaled to
different values so long as the ionization parameter remains the same,
which will guarantee that the ionization fractions in the model are
not affected by the rescaling. This is because the dimensionless form
of the gas-dynamical, radiation transfer and ionization equations
contain only a handful of dimensionless parameters \citep[for
example][ Appendix~A]{2005ApJ...621..328H}. The only one of these that
depends on the density and length scales is a dimensionless form of
the ionization parameter: $\tau_* = n_0 \sigma_0 \ell_0$, where
$\ell_0$ is the length scale.

One can then use the Str\"omgren relation, $Q\Sub{H} \propto n_0^2
\ell_0^3$, to eliminate the density\footnote{Although ionizations
  exceed recombinations by a small ($\simeq 10/\tau_*$) fraction, this
  fraction is constant for all models with the same $\tau_*$.} and find
that a fixed ionization parameter requires a fixed ratio of
$Q\Sub{H}/\ell_0$. All times should also be scaled proportionately to
$\ell$.  Furthermore, it is even possible to scale $Q\Sub{H}$ and
$\ell$ independently, subject to the understanding that one is also
thereby scaling the effective photoionization cross-section as
$\sigma_0 \propto (\ell/Q\Sub{H})^{1/2}$.

\section{Model Results}
\label{sec:model-results}

\subsection{Early Evolution of the  Models}
\label{sec:early-model-evolution}

\begin{figure}
\includegraphics{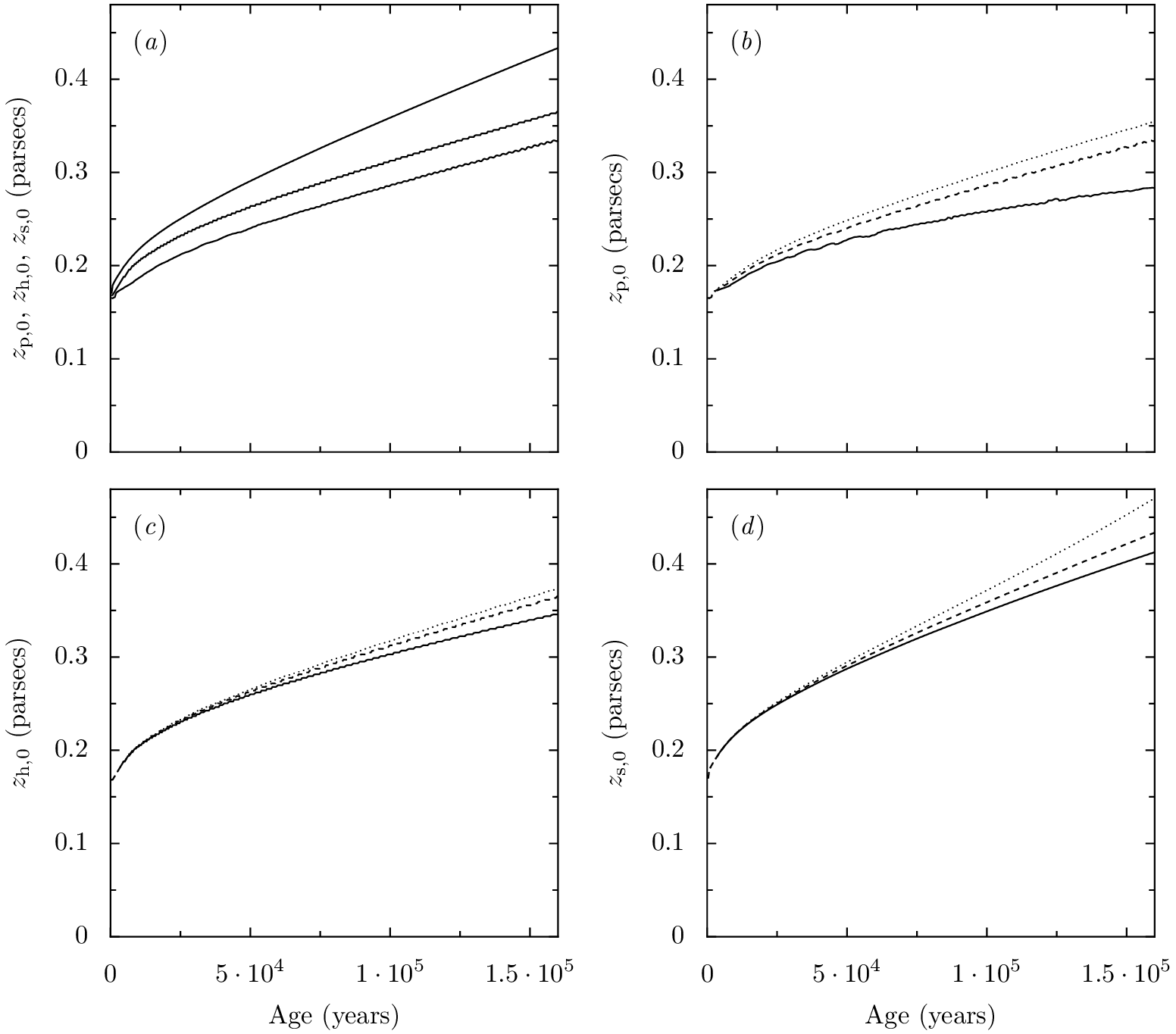}
 \caption{(\textit{a})~Evolution with time of the axial distance from
    the star to the shock front, $z_\mathrm{s}$ (upper line), heating
    front $z_\mathrm{h}$ (middle line), and ionized density peak,
    $z_\mathrm{p}$ (lower line), all for Model~B.
    (\textit{b})~Evolution of $z_\mathrm{p}$ for the different models:
    A (solid line), B (dashed line), and C (dotted line).
    (\textit{c})~Same as \textit{b} but for the evolution of
    $z_\mathrm{h}$. (\textit{d})~Same as \textit{b} but for the
    evolution of $z_\mathrm{s}$.}
  \label{fig:evo-z}
\end{figure}

The initial evolution of all three models is the same: the low-density
gas below the interface is completely ionized on a recombination
timescale of $\sim 1000\,\years$. At the interface itself, the
ionization/heating front at first penetrates to gas of density $\sim
\rho_0$, which is heated to $\simeq 9000\,\Kelvin$, and achieves a
thermal pressure 30 times higher than the pressure in the cold neutral
gas. This drives a shockwave up the density ramp at a speed $u\Sub{s}
\simeq (\rho\Sub{i}/\rho\Sub{n})^{1/2} c\Sub{i}$, where $\rho\Sub{i}$
is the ionized density, $c\Sub{i}$ is the ionized sound speed, and
$\rho\Sub{n}$ is the pre-shock neutral density.\footnote{The shock
  actually goes a bit faster than this equation suggests since the
  pressure at the heating front (see below) is about twice the
  pressure in the fully ionized gas.} The shock can be first discerned
in the simulations after about $3000\,\years$, at which time its speed
is $u\Sub{s} \simeq 5\,\kms$, falling to $\simeq 1\,\kms$ by the time
it reaches the top of the density ramp (after about $40,000\,\years$).
The time-evolution of the shock front position in the three models is
almost identical during this early stage, as can be seen in
Figure~\ref{fig:evo-z}\textit{d}.

At the same time as the shock propagates into the neutral gas, the
warm gas expands down the density ramp, towards (and beyond) the star.
Within $30,000\,\years$ all the gas that was originally below the
density interface has been swept off the grid by this flow, which
accelerates up to about $20 \,\kms$. Initially, the ionized gas flows
parallel to the $z$-axis but as time goes on the streamlines begin to
diverge and the flow settles into a quasi-steady configuration, during
which the properties of the ionized flow and the subsequent evolution
of the ionization front vary significantly between the models.

In addition to the position of the shock front, $z_\mathrm{s}$, we
make reference below to the position of the heating front,
$z_\mathrm{h}$, which is a very thin layer at the very base of the
evaporation flow, and also to the position of the ionized density
peak, $z_\mathrm{p}$, formed by the opposing gradients in total
density and ionization fraction. The time evolution of these three
quantities is shown in Figure~\ref{fig:evo-z}.

\subsection{Properties of the Quasi-steady Flow}
\label{sec:quasi-steady}

\begin{figure*}
  \includegraphics{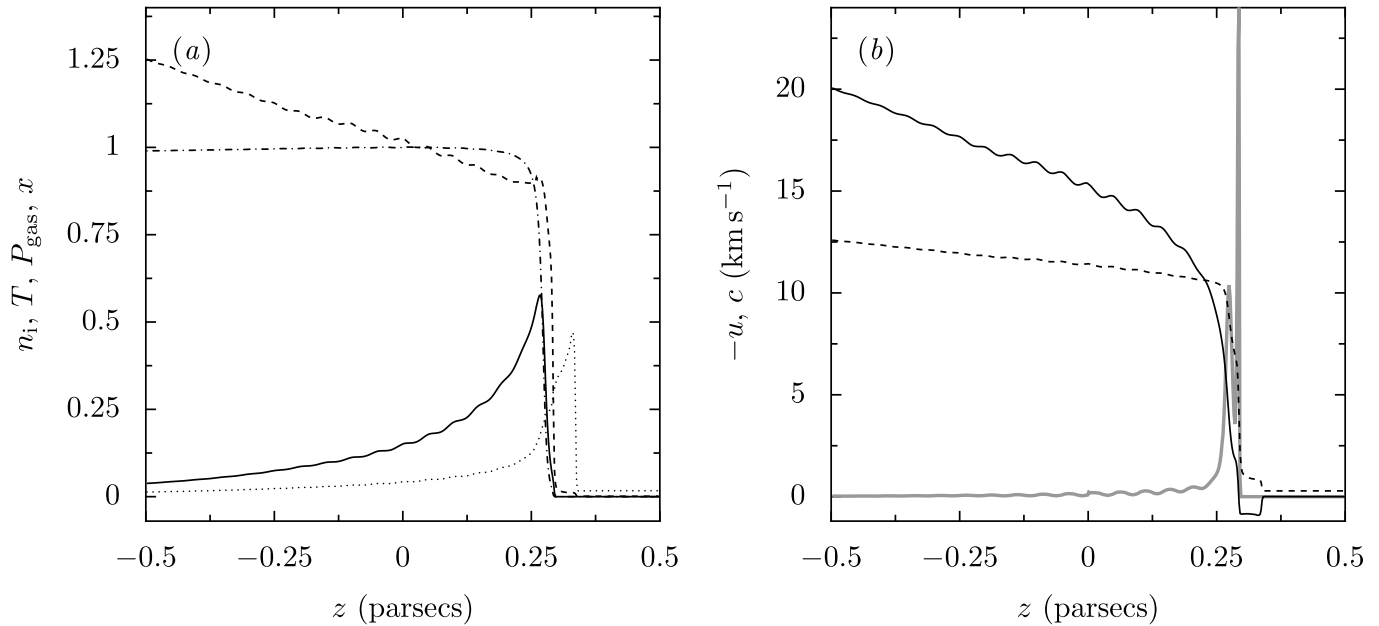}
  \caption{Axial profiles of physical variables in Model~B at an age
    of $78,000\,\years$. (\textit{a})~Ionized hydrogen density
    $n_\mathrm{i}$ (solid line, units of $10^4\,\pcc$), gas
    temperature $T$ (dashed line, units of $10^4\,\Kelvin$), pressure
    $P_\mathrm{gas}$ (dotted line, units of $10^{-7}\,\punits$), and
    ionization fraction (dot-dashed line). (\textit{b})~Negative of
    the gas velocity in the reference frame of the star, $u$ (solid
    line), and isothermal sound speed (dashed line, both in $\kms$),
    plus net heating rate (thick gray line, arbitrary units).}
  \label{fig:axial}
\end{figure*}

\begin{figure*}
  \includegraphics{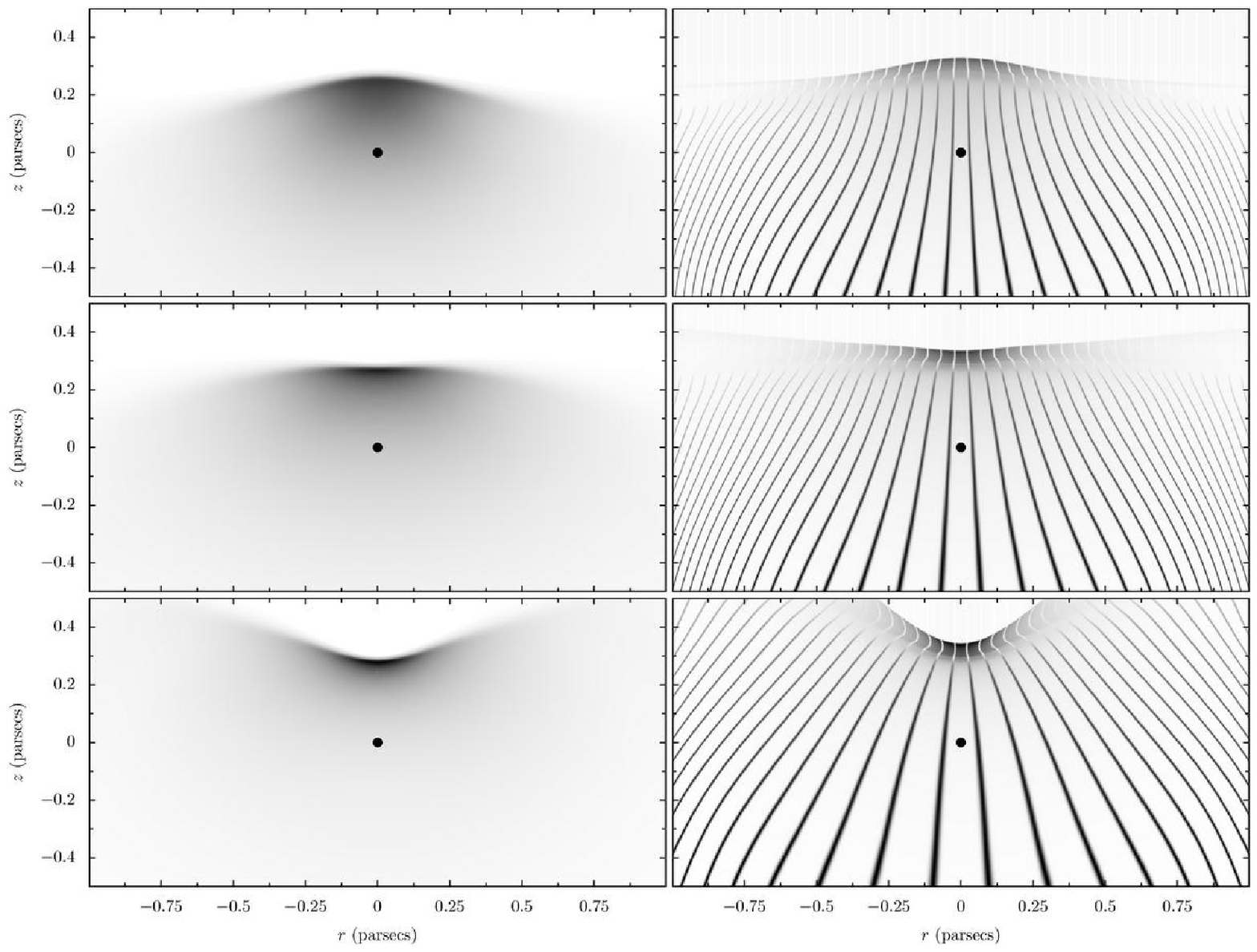}
  \caption{Snapshot images of the three models (top to bottom: A, B,
    C) at a time of $78,000\,\years$.  Left column shows ionized
    hydrogen density as a linear negative grayscale, where black
    represents a density of $6600\,\pcc$. Right column shows gas
    pressure ($\mathrm{black} = 6.1\times 10^{-8}\,\punits$) with
    superimposed streamlines whose darkness indicates magnitude of the
    gas velocity ($\mathrm{black} = 20.1\,\kms$).  All grayscales are
    common between the models. The position of the ionizing star is
    indicated by a filled circle in each panel. }
  \label{fig:snapshot}
\end{figure*}

After the initial transient phase, the flow settles down into a long
quasi-steady phase in which the flow properties change only on
timescales much longer than the dynamic time for flow away from the
ionization front. As we will show below, the flow structure, once
normalized to the distance of the star from the front, seems to depend
only on the curvature of the ionization front. During this
quasi-steady phase the shock in the neutral gas is propagating at
about $1\ \kms$, as discussed in the previous section, and the
post-shock gas moves at about three-quarters of this speed since it
does does not cool strongly in our simulations. At the same time, the
relative velocity between the ionization/heating front and the neutral
gas immediately outside it is very low (for a \hbox{D-critical} front this
would be $c\Sub{n}/2c\Sub{i}^2 \simeq 0.05\ \kms$, see
Appendix~\ref{sec:class-ioniz-front}), so that the ionization front
speed is also about three-quarters of the shock speed, leading to a
steady thickening of the shocked neutral layer.  However, this
behavior is dependent on the cooling behavior of the shocked neutral
gas, which we do not treat very realistically.  Stronger cooling would
cause the ionization and shock fronts to propagate at the same speed.

Details of the variation along the $z$ axis of various physical
quantities for Model~B are shown in Figure~\ref{fig:axial}.  Images of
the distributions of ionized density and pressure in the models at a
representative point in their evolution are shown in
Figure~\ref{fig:snapshot}, together with the flow streamlines.

\subsubsection{Structure along the $z$-axis}
\label{sec:structure-along-z}

The shock driven into the neutral gas is traced by the pressure
(dotted line in Fig.~\ref{fig:axial}\textit{a}, grayscale in
right-hand panels of Fig.~\ref{fig:snapshot}), which peaks immediately
behind the shock.  The ionization front is traced by the ionization
fraction (dot-dashed line in Fig.~\ref{fig:axial}\textit{a}) and the
ionized density (solid line in Fig.~\ref{fig:axial}\textit{a}) shows a
peak just inside the front. One can also see a heating front, best
seen in the temperature (dashed line in
Fig.~\ref{fig:axial}\textit{a}), which lies just outside the
ionization front. The temperature shows a slight peak at the
ionization front due to the radiation hardening.

\begin{figure}\centering
 \includegraphics{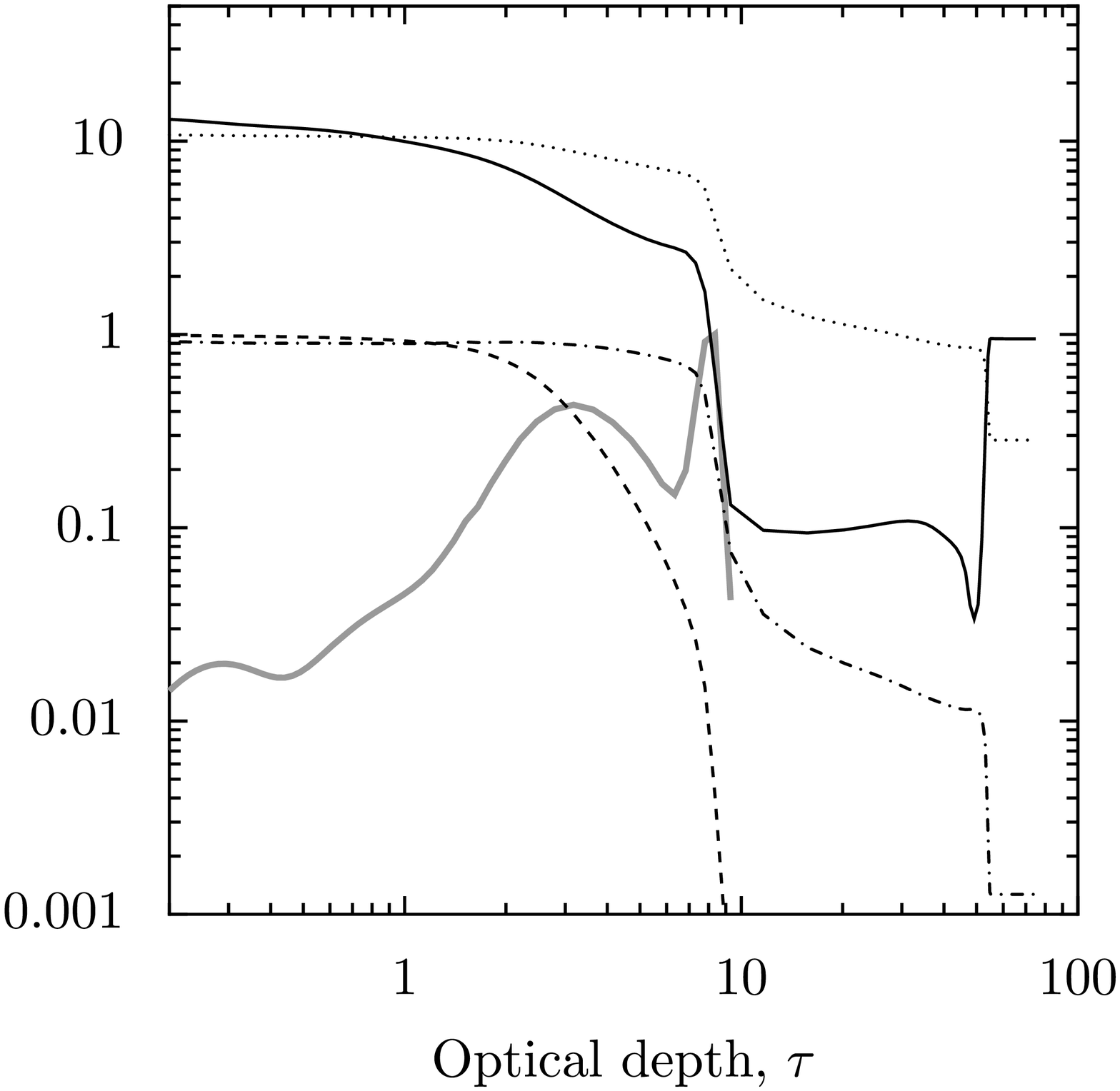}
 \caption[]{Structure of the ionization front in Model~B at an age of
   $78,000\,\years$. Gas velocity with respect to the heating front,
  $v$, in $\kms{}$ (solid line); ionization fraction, $x$ (dashed
  line); isothermal sound speed, $c$, in $\kms{}$ (dotted line);
  temperature, $T/10^4\,\Kelvin$ (dot-dashed line); net energy
  deposition rate in arbitrary units (thick gray line).}
  \label{fig:axial-x}
\end{figure}

The solid line in Figure~\ref{fig:axial} shows the negative of the gas
velocity (i.e. velocity back towards the ionizing star). The neutral
gas is initially accelerated away from the star by the shock, which
gives the neutral shell a speed of $\sim +1 \,\kms$. In the ionization
front it is accelerated back towards the star, quickly reaching speeds
of $\sim -10 \,\kms$ and then continuing a gradual acceleration to
$\sim -20 \,\kms$ before leaving the grid. The isothermal sound speed
is shown by the dashed line in Figure~\ref{fig:axial}\textit{b} and it
can be seen that the flow becomes supersonic shortly after passing the
peak in the ionized density.  The question of whether the fronts are
\hbox{D-critical} is addressed in
Appendix~\ref{sec:class-ioniz-front}.
The net rate of energy deposition into the gas (photoionization
heating minus collisional line cooling) is shown by the thick gray
line in Figure~\ref{fig:axial}\textit{b} and always shows two peaks.
These can be seen in more detail in Figure~\ref{fig:axial-x}, which
shows a log-log plot of various quantities as a function of the
optical depth to ionizing radiation, $\tau = \int \langle\sigma\rangle
n(\mathrm{H_0})\, d z$.  The deeper of the two energy deposition peaks
is the heating front, which occurs at an ionizing optical depth of
$\tau \simeq 8$, where the ionization fraction is around $x \simeq
0.01$ and the gas density around $n \simeq 10^{5} \, \pcc$. In this
front the gas is rapidly heated up to just under $10^4\,\Kelvin$ and
accelerated up to $2$--$3\,\kms$ but remains largely neutral (note
that in Fig.~\ref{fig:axial-x} the velocity is now shown in the rest
frame of the heating front, $v$).  The second, broader peak, at $\tau
= 2$--$4$, coincides with the ionization of the gas at a roughly
constant temperature and its gradual acceleration up to the ionized
sound speed.

\subsubsection{Differences between the models}
\label{sec:later-struct-models}

Despite the general similarities, there are some striking differences
between the physical structure of the three models, as can be seen
from Figure~\ref{fig:snapshot}. In Model~A ($\alpha=0$), where the
neutral gas has a constant density in the $r$-direction, both the
shock front and the ionization front are concave from the point of
view of the star, whereas in Model~B ($\alpha=1$), where the neutral
gas density falls off with radius aymptotically as $r^{-2}$, the
fronts are approximately flat. An analytic treatment of the flow in
this special case is given in Appendix~\ref{sec:analytic-model-flow}.
In Model~C ($\alpha=2$), which has an even steeper radial density
profile, the fronts are convex from the point of view of the star. As
a result of these changes, one sees several clear trends going from
Model~A, to~B, to~C---the ionized density falls off faster in the
direction towards the star; the ionized flow accelerates more sharply;
the flow streamlines show greater divergence; the pressure driving the
shock is greater; and, the ionization front progresses faster at late
times.

\begin{figure}\centering
  \includegraphics{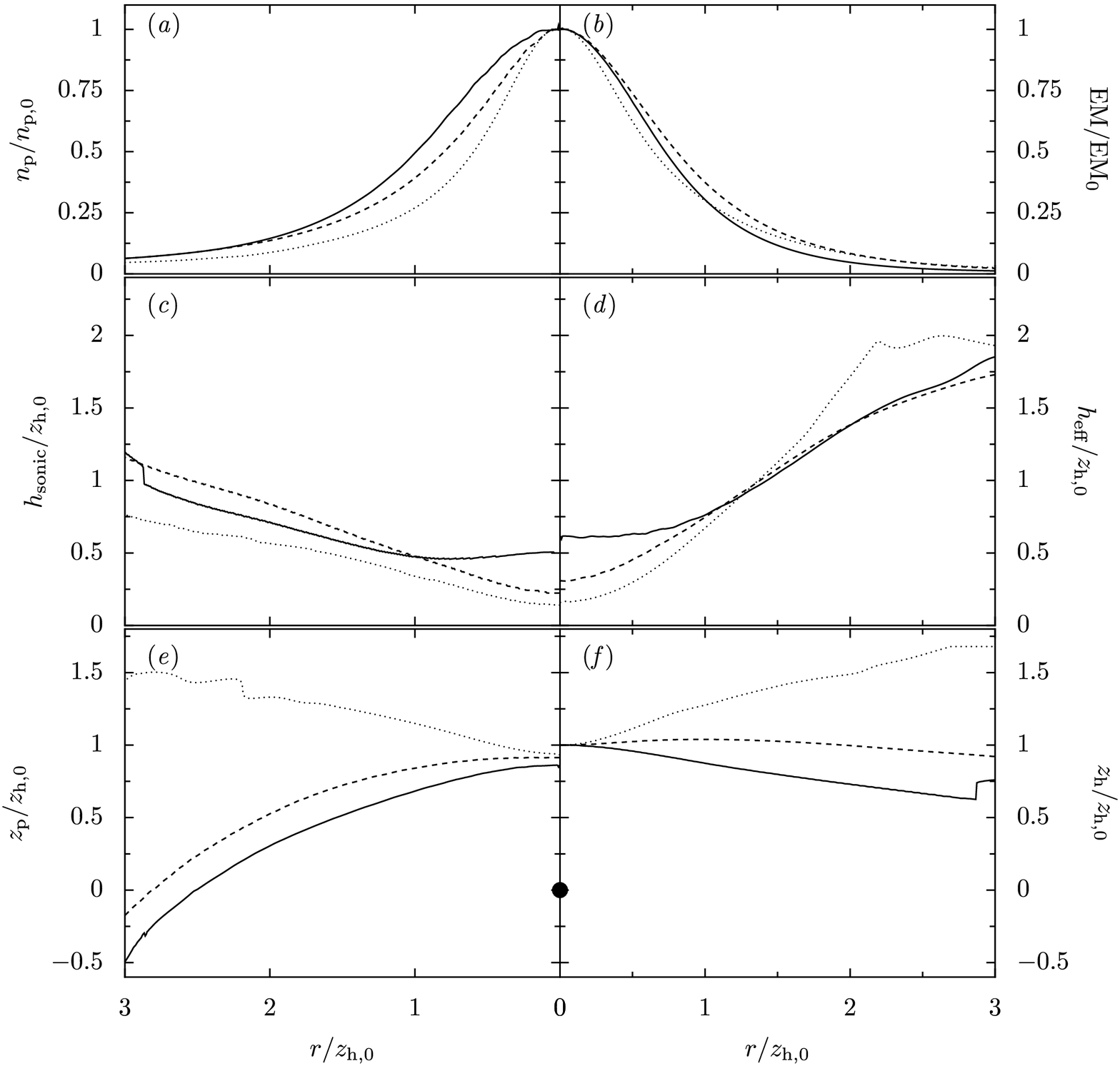}
  \caption{Lateral variation of various model parameters at an age of
    $78,000\,\years$. All lengths are normalized to $z_\mathrm{h,0}$,
    the distance between the star and the heating front along the
    axis, and are plotted against cylindrical radius. The different
    models are shown by: A (solid line), B (dashed line), and C
    (dotted line).  (\textit{a})~Peak ionized density at each radius,
    normalized to value on the axis.  (\textit{b})~Total ionized
    emission measure, integrated along the $z$-direction: $\mathrm{EM}
    = \int n_\mathrm{i}^2 \, dz$.  (\textit{c})~Sonic thickness,
    $h_\mathrm{sonic}$, defined as the $z$-distance between the
    heating front and the sonic point.  (\textit{d})~Effective
    recombination thickness, $h_\mathrm{eff} = \mathrm{EM}/
    n_\mathrm{p}^2$.  (\textit{e})~Position of the ionized density
    peak, $z_\mathrm{p}$.  (\textit{f})~Position of the heating front,
    $z_\mathrm{h}$.}
  \label{fig:lateral}
\end{figure}

The lateral structure (Figure~\ref{fig:lateral}) is roughly similar
for all 3 models, except for the electron density, which falls off
more rapidly in Model~C, and the flow thickness close to the axis,
which is significantly larger for Model~A\@.

\subsection{Secular Evolution of the Flow in the Quasi-Steady Regime}
\label{sec:secul-evol-flow}

\begin{figure}\centering
  \includegraphics{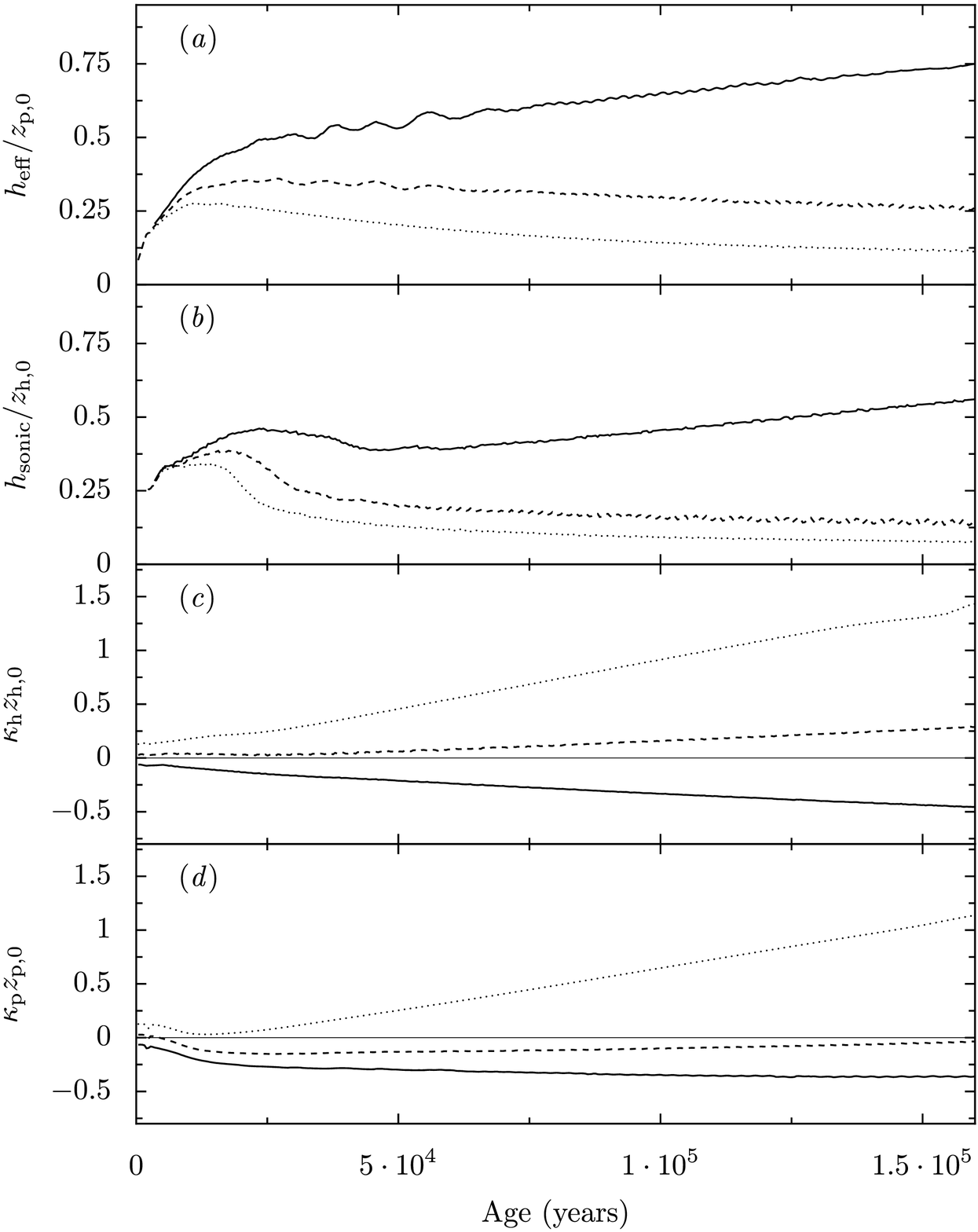}
  \caption{Evolution with time of the flow thickness along the
    $z$-axis and curvature, normalised to the distance,
    $z_\mathrm{h,0}$, between the star and the heating front.
    (\textit{a})~Effective recombination thickness. (\textit{b})~Sonic
    thickness. (\textit{c})~Curvature of heating front.
    (\textit{d})~Curvature of ionized density ridge. The different
    models are shown by: A (solid line), B (dashed line), and C
    (dotted line). The thin horizontal line in the bottom two panels
    indicates zero curvature and is the dividing line between convex
    and concave surfaces.}
  \label{fig:evo-h}
\end{figure}

Figure~\ref{fig:evo-h} shows how the effective thickness (panel
\textit{a}) and sonic thickness (panel \textit{b}) of the flows vary
with time for the different models. It can be seen that once the
quasi-steady regime is reached ($t > 5\times10^4\,\years$), the flow
thickness increases with time for Model~A, but decreases with time for
Model~C, with Model~B showing a gentler decrease.

The curvature $\kappa$ of each surface $z(r)$ was found by fitting a
parabola, $z = z_0 + 0.5\kappa r^2$, to the surface for
$r<0.25\,\parsec$. Thus $\left|\kappa\right|^{-1}$ is equal to the
radius of curvature of the surface, with positive $\kappa$
corresponding to a convex surface that curves away from the ionizing
star and negative $\kappa$ corresponding to a concave surface that
curves towards the ionizing star. Figures~\ref{fig:evo-h}\textit{c} and
\textit{d} show the evolution with time of the dimensionless
curvature, $\kappa z_0$, calculated for the heating front and for
the maximum in the ionized density, respectively.

\begin{figure}\centering
  \includegraphics{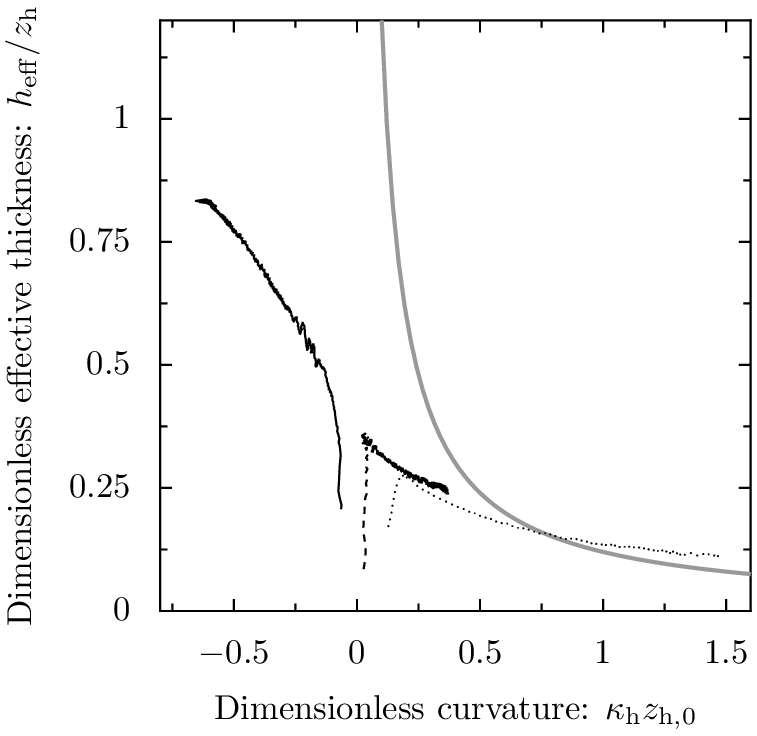}
  \caption{Effective thickness of the photoevaporation flow on the
    $z$-axis versus heating front curvature. Lines represent models A
    (solid), B (dashed), and C (dotted).  The thick gray line shows
    the expected relationship for a ``globule'' flow, $h =
    0.12/\kappa$. }
  \label{fig:evo-kappa}
\end{figure}

Figure~\ref{fig:evo-kappa} graphs the dimensionless effective
thickness against dimensionless curvature for all models and all
times. It can be seen that the vast majority of points approximately
follow a single curve for the three models.\footnote{The points that
  lie below this curve, at small values of $\kappa$, correspond to the
  initial non-steady evolution at the beginning of each model run.}
This is consistent with the idea that the flow thickness is controlled
by the curvature of the ionization front. When the front is concave
($\kappa_\mathrm{h} < 0$), the flow is not very divergent until it has
passed the ionizing star, leading to a flow thickness of order the
star-front distance. When one moves to flat and convex fronts, the
flow divergence becomes more pronounced and its thickness becomes
considerably less than the star-front distance. In the limit of large
convex curvature ($\kappa_\mathrm{h} z_\mathrm{h,0} \gg 1$), then one
would expect the flow to resemble that from a cometary globule
\citep{1990ApJ...354..529B}, with the flow thickness proportional to
the radius of curvature: $h = \omega/\kappa$, where $\omega \simeq
0.12$. This curve is shown by the thick gray line in
Figure~\ref{fig:evo-kappa} and it can be seen that our models are
still some way from this regime.

\subsection{Optical Line Emission}
\label{sec:optic-line-emiss}

We have calculated the emissivity for three different types of line,
representative of the strongest optical lines emitted by \hii{}
regions. The first is a recombination line, such as the hydrogen
Balmer lines, with emissivity
\begin{equation}
  \label{eq:emissivity-recombination}
  \eta_\mathrm{rec}(x) \propto n_\mathrm{e} n_{j+1} \, T^{-1},
\end{equation}
where $n_{j+1}$ is the number density of the recombining ion, which we
take to be proportional to $n_\mathrm{i}$. The second and third are
both forbidden metal lines, excited by electron collisions, with
emissivity 
\begin{equation}
  \label{eq:emissivity-collisional}
  \eta_\mathrm{col}(x)  \propto \frac{n_\mathrm{e}
    n_j}{\displaystyle 1 + B_\mathrm{col} n_\mathrm{e} T^{-1/2}} \, 
  \frac{e^{-E/kT}}{ \sqrt{T} } , 
\end{equation}
where $E$ is the excitation energy of the upper level, which we fix at
$h c / 6500\,\Angstrom$, and $B_\mathrm{col}$ is the collisional
de-excitation coefficient, chosen to give a critical density of
$1000\,\pcc$ at $10,000\,\Kelvin$.  The second line is emitted by a
neutral metal with $n_j \propto n_\mathrm{n}$, whereas the third line
comes from an ionized metal with $n_j \propto n_\mathrm{i}$. Although
the emissivities that we use are generic, they are each ``inspired''
by a particular optical emission line. In particular, the
recombination line represents a hydrogen Balmer line such as
H$\alpha$, the collisional neutral line represents \OIlam{}, and the
collisional ionized line represents a combination of \NIIlam{} and
\OIIIlam{} (since we do not consider the ionization of helium in our
models, we cannot distinguish between singly and doubly ionized
metals).

\begin{figure}\centering
  \ifpdf 
    \includegraphics[angle=90,width=\linewidth]{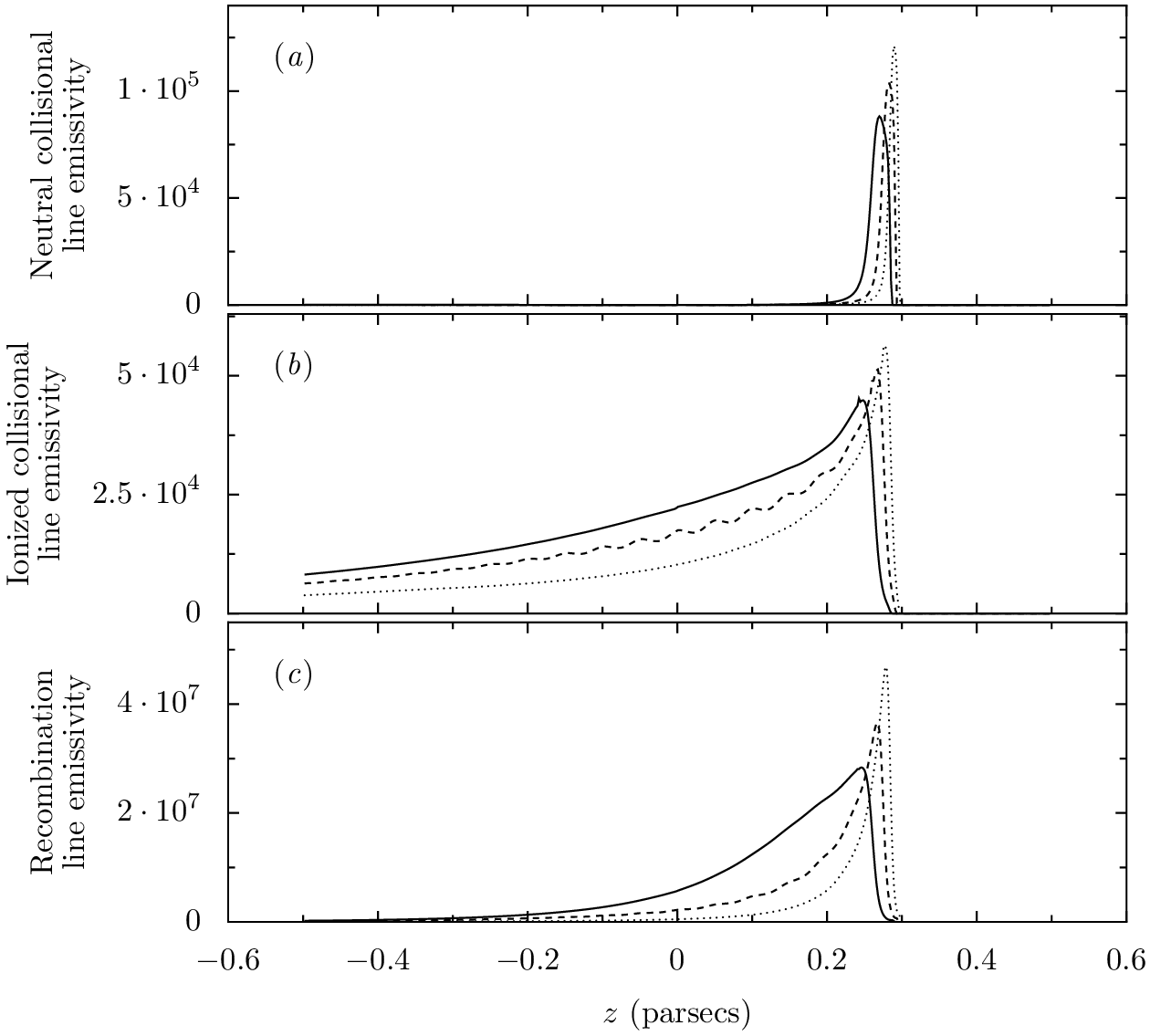}
  \else
    \includegraphics{f08}
  \fi
  \caption{Axial profiles of emissivity in different emission lines at
    an age of $78,000\,\years$. The different models are shown by: A
    (solid line), B (dashed line), and C (dotted line).
    (\textit{a})~Collisionally excited neutral metal line.
    (\textit{b})~Collisionally excited ionized metal line.
    (\textit{c})~Recombination line. }
  \label{fig:axial-emiss}
\end{figure}

Figure~\ref{fig:axial-emiss} shows the emissivity of each line as a
function of $z$ for each model, with arbitrary normalization. It can
be seen that the neutral collisional line is confined to a thin region
around the ionization front. This is because its emissivity contains a
factor of $n_\mathrm{e} n_\mathrm{n} \propto x (1-x)$, which peaks at
$x=0.5$. The other two lines show emissivity profiles that are broadly
similar to one another since they are both proportional to
$n_\mathrm{i}^2$. The differences between these two are mainly due to
their different temperature dependences: the recombination line
emissivity is relatively stronger at lower temperatures near the
ionization front, while the collisional line is relatively stronger at
higher temperatures, farther out in the flow. This behavior is further
accentuated by the collisional deexcitation of the metal line, which
reduces its strength at the higher densities found close to the
ionization front.

\begin{figure}\centering
  \ifpdf 
    \includegraphics[angle=90,width=\linewidth]{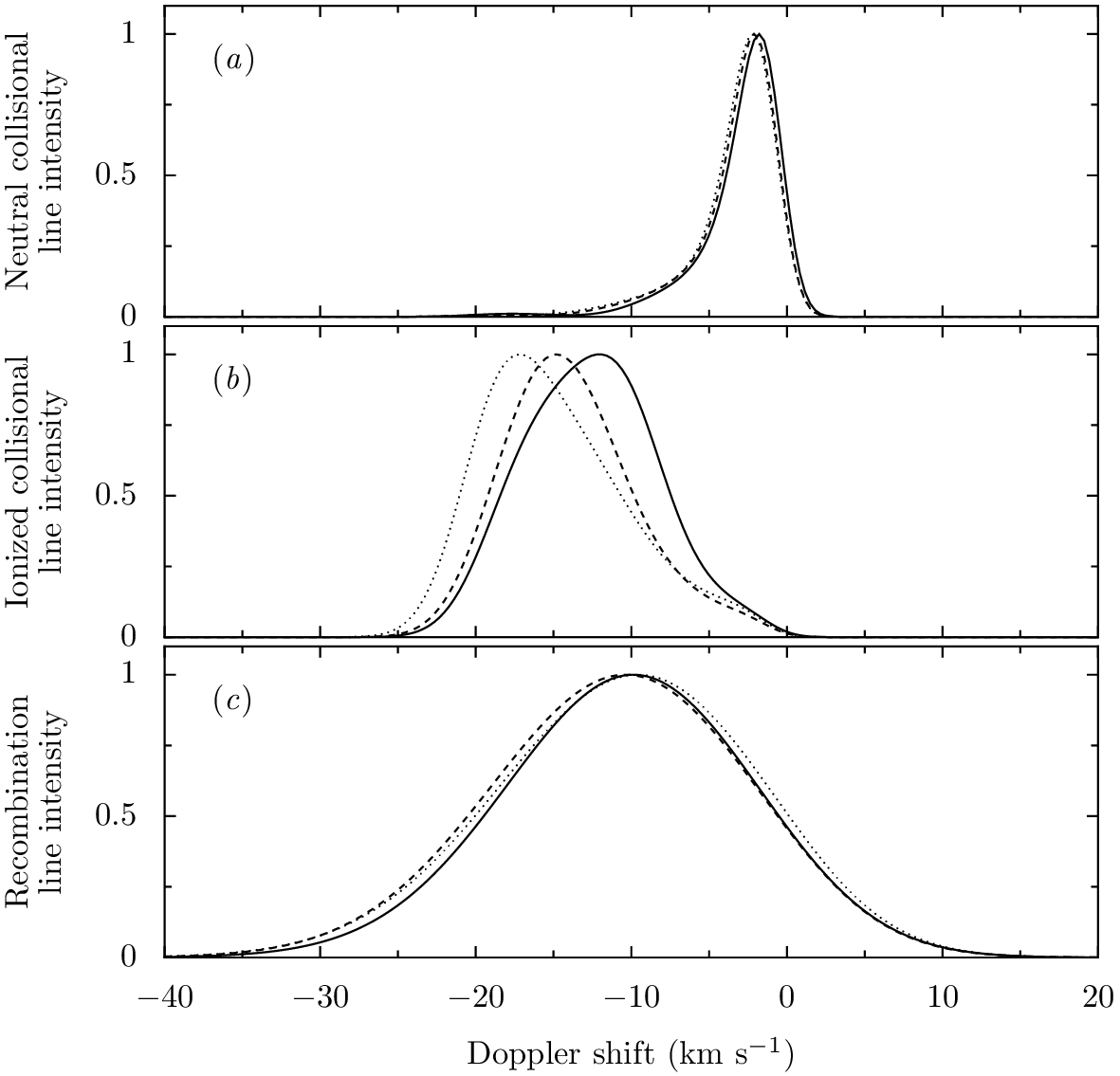}
  \else
    \includegraphics{f09}
  \fi
  \caption{Simulated emission line profiles for a face-on viewing
    angle at an age of $78,000\,\years$. The different models are
    shown by: A (solid line), B (dashed line), and C (dotted line).
    (\textit{a})~Collisionally excited neutral metal line with atomic
    weight $A=16$. (\textit{b})~Collisionally excited ionized metal
    line with $A=14$. (\textit{c})~Recombination line with $A=1$.}
  \label{fig:vel-profiles}
\end{figure}

We have also calculated the spectroscopic emission line profiles that
would be seen by an observer looking along the $z$-axis. We assume
that the radiative transfer can be calculated in the optically thin
limit, in which case the emergent intensity profile is
\begin{equation}
  \label{eq:line-profile}
  I(u) = 
\int_0^\infty \eta(z) \, \exp\left[ -\frac{ \left( v(z) - u
      \right)^2 }{ 2\Delta^2(z) } \right] \, dz , 
\end{equation}
where $u$ is the observed Doppler shift and $\Delta(z)$ is the thermal
width at each point in the structure, which depends on the local sound
speed and on the atomic weight, $A$, of the emitting species. The
results are shown in Figure~\ref{fig:vel-profiles}, assuming $A=1$ for
the recombination line, $A=14$ for the ionized metal line, and $A=16$
for the neutral metal line.

The neutral metal line shows only a small blueshift ($\simeq 2\,
\kms$) of its peak with respect to the quiescent gas, but also has an
extended blue wing. Its profile is almost indistinguishable between
the different models. The ionized metal line shows the greatest
variation between models but all are qualitatively similar with a
large net blueshift (12--$17\,\kms$) and a FWHM of $\simeq 10\,\kms$,
which is roughly double the thermal width. In cases in which the
viewing angle has been determined by independent means, this line may
be used to discriminate kinematically between the different models.
The recombination line is also blueshifted (by about $10\,\kms$) and
again shows only slight variation between the models, with a FWHM that
is now dominated by the thermal width because of the lower mass of H.  

\begin{figure}\centering
  \includegraphics{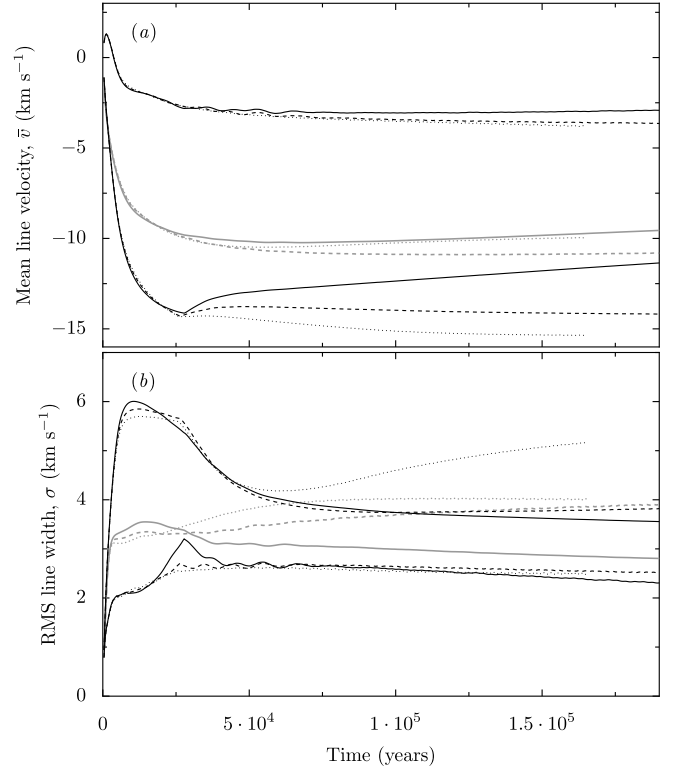}
  \caption{Evolution of emission line kinematic properties for a
    face-on viewing angle for
    the three models. The different models are shown by: A (solid
    line), B (dashed line), and C (dotted line). (\textit{a})~Mean
    line velocity, $\bar{v}$. (\textit{b})~Root-mean-squared line
    width, $\sigma$. In panel (\textit{a}), the top set of lines are
    for the neutral collisional line, the middle gray set are for the
    recombination line, and the bottom set are for the ionized
    collisional line. In the panel (\textit{b}) the order is
    reversed.}
  \label{fig:evo-lines}
\end{figure}

The variation in the properties of the synthesized line profiles as
the models evolve is presented in Figure~\ref{fig:evo-lines}, which
shows the mean velocity,
\begin{equation}
  \label{eq:mean-velocity}
  \bar{v} 
  = \dfrac{\int_0^\infty v(z) \eta(z) \,dz}{\int_0^\infty \eta(z)
    \,dz} , 
\end{equation}
and the RMS velocity width, 
\begin{equation}
  \label{eq:rms-line-width}
  \sigma 
  = \dfrac{\int_0^\infty (v(z)-\bar{v})^2 \eta(z) \,dz}{\int_0^\infty \eta(z)
    \,dz} ,
\end{equation}
calculated for the three models as a function of time. Note that the
velocity width, $\sigma$, only includes the contribution from the
acceleration of the gas and does not include the thermal Doppler
broadening. Supposing the emission to come predominantly from gas at
$10^4\,\Kelvin$, the observed FWHM should be related to the RMS width
by $\mathrm{FWHM} \simeq [(20\,\kms/A)^2 + (2.355 \sigma)^2]^{1/2} $.

Once the models have settled down to their quasi-steady stage of
evolution, it is remarkable how little variation is seen in the
emission line properties, either between the models or as a function
of time.

\section{Observational parameters of the Orion Nebula}
\label{sec:observational-data}

In this section, we derive observational parameters for the Orion
nebula in order to compare with our simulations. We choose to use the
spatial variation of the emission measure and electron density,
together with the kinematic properties of various optical emission lines.

\subsection{Emission measure and electron density}
\label{sec:emiss-meas-electr}
\begin{figure*}\centering
  \includegraphics{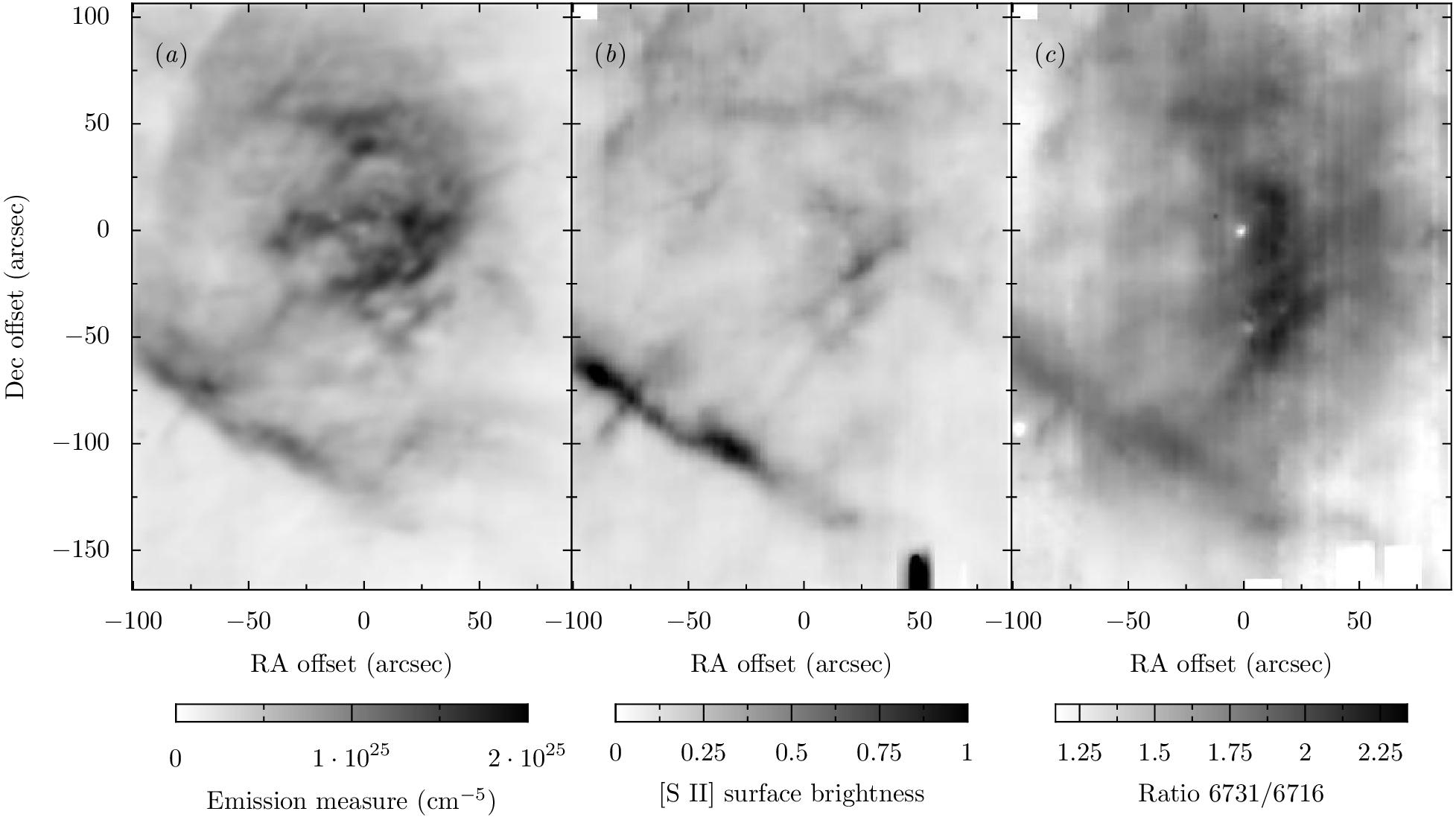}
  \caption{(\textit{a})~Emission measure. (\textit{b})~[\ion{S}{2}]
    6716+6731\,\AA{} surface brightness. (\textit{c})~[\ion{S}{2}]
    $6731/6716$ ratio. }
  \label{fig:orion}
\end{figure*}
Emission measure and electron density maps of the inner Orion nebula
are shown in Figure~\ref{fig:orion}.  We base these maps on
velocity-resolved data cubes of the emissivity in various optical
emission lines \citep{garcia03,2004Doi-kinematics}.  This allows us to
distinguish between the emission from the photoevaporation flow and
that produced by Herbig-Haro shocks \citep{1997AJ....114..730O} and
back-scattering from dust in the molecular cloud
\citep{1998ApJ...503..760H}.

The emission measure map is calculated from the H$\alpha$ data of
\citet{2004Doi-kinematics}, using a heliocentric velocity interval of
$+1$ to $+33\,\kms$, and correcting for foreground dust extinction
using the $C(\mathrm{H\beta})$ map of \citet{2000AJ....120..382O} and
the extinction curve of \citet{1989ApJ...345..245C}. The
emission measure is assumed to be strictly proportional to the
H$\alpha$ surface brightness, which would be the case for isothermal,
Case~B emission, and the constant of proportionality is found by
comparison with the measured free-free optical depth at 20\,cm
\citep{2000AJ....120..382O}, assuming an electron temperature of
$8900\,\Kelvin$\@. In regions of the map that do not contain
significant high-velocity emission our derived emission measure is
identical to that derived directly from the radio free-free data.

As an additional check on our results, we flux-calibrated our
H$\alpha$ map (using the full velocity range) by comparison with
published \textit{HST} WFPC2 imaging
\citep{1999PASP..111.1316O,2000AJ....119.2919B} and converted to
emission measure using the atomic parameters in
\citet{Osterbrock-book}, which yielded values consistent at the 10\%
level with those derived by \citet{1991ApJ...374..580B} from the
H\,11--3 line. As recognised previously \citep{1995ApJ...438..784W},
the emission measures derived from the full velocity range of optical
recombination lines are overestimated by up to 50\% because of the
contribution of light that has been back-scattered by dust in the
molecular cloud. Our velocity-resolved technique automatically
compensates for this without the need to introduce an arbitrary
correction factor.

In order to calculate the electron density in the photoevaporation
flow, we use velocity-resolved maps of the density-sensitive doublet
lines [\ion{S}{2}]\,6716 and 6731\,\AA{}
\citep{garcia03,GarciaHenney2005} to select only that gas in the range
$+4$ to $+24\,\kms$, close to the peak velocities of fully ionized
species.
The summed surface brightness in this velocity
range for the two lines is shown in Figure~\ref{fig:orion}\textit{b}
after correction for foreground extinction (the apparent emission peak
at the lower right of the image is an artifact of the reduction
procedure). The density-sensitive ratio $R = 6731/6716$ for the same
velocity range is shown in Figure~\ref{fig:orion}\textit{c}, in which
higher values of $R$ correspond to higher densities. In order to
calibrate our ratios, we first calculated maps summed over the full
velocity range of the [\ion{S}{2}] emission and compared these with
the spectrophotometric data of \citet{1991ApJ...374..580B} and
\citet{1992ApJ...399..147P}.

We find that a very good fit to the dependence of electron density on
$R$ for gas at $8900\,\Kelvin$ \citep{1993ApJS...88..329C} is
\begin{equation}
  \label{eq:eden-ratio}
  n_\mathrm{e} = n_0 \frac{R-R_1}{R_2-R}, 
\end{equation}
where $n_0 = 2489\,\pcc$, $R_1 = 0.697$, and $R_2 = 2.338$. Using this
fit, the smallest ratio shown in Figure~\ref{fig:orion}\textit{c}
corresponds to $n_\mathrm{e} \simeq 1000\,\pcc$ and the mid-range
corresponds to 4--$6000\,\pcc$. The largest ratios shown in the image
are in the high-density limit ($n_\mathrm{e} > 10^4\,\pcc$), at which
point the derived density becomes very uncertain due to the extreme
sensitivity to experimental errors and variations in the temperature.
Previous work, which did not resolve the kinematic line profiles,
found that electron densities derived from fully ionized species
\citep{1992PhDT........35J,1990ApJ...356..534M} are
generally higher than those determined from [\ion{S}{2}]. This has
been attributed to the fact that part of the [\ion{S}{2}] emission
comes from partially ionized zones in which the electron fraction is
significantly less than unity. As a result, it was found necessary to
apply a correction factor to the [\ion{S}{2}] densities, although
there has been significant disagreement on its value
\citep{1991ApJ...374..580B,1995ApJ...438..784W}. By using only the
blue-shifted portion of the line profiles to calculate our densities,
we are selecting the [\ion{S}{2}] emission from the fully ionized gas.
As a result, we have no need to apply a correction factor to our
densities. Indeed we find that the $6731/6716$ ratio for the
blue-shifted velocity range is typically 10\% larger than the ratio
calculated for the entire line, although there are some local
variations and the difference is less at greater projected distances
from the ionizing star.

\subsection{Gas kinematics}
\label{sec:gas-kinematics}
\begin{figure*}\centering
  \includegraphics{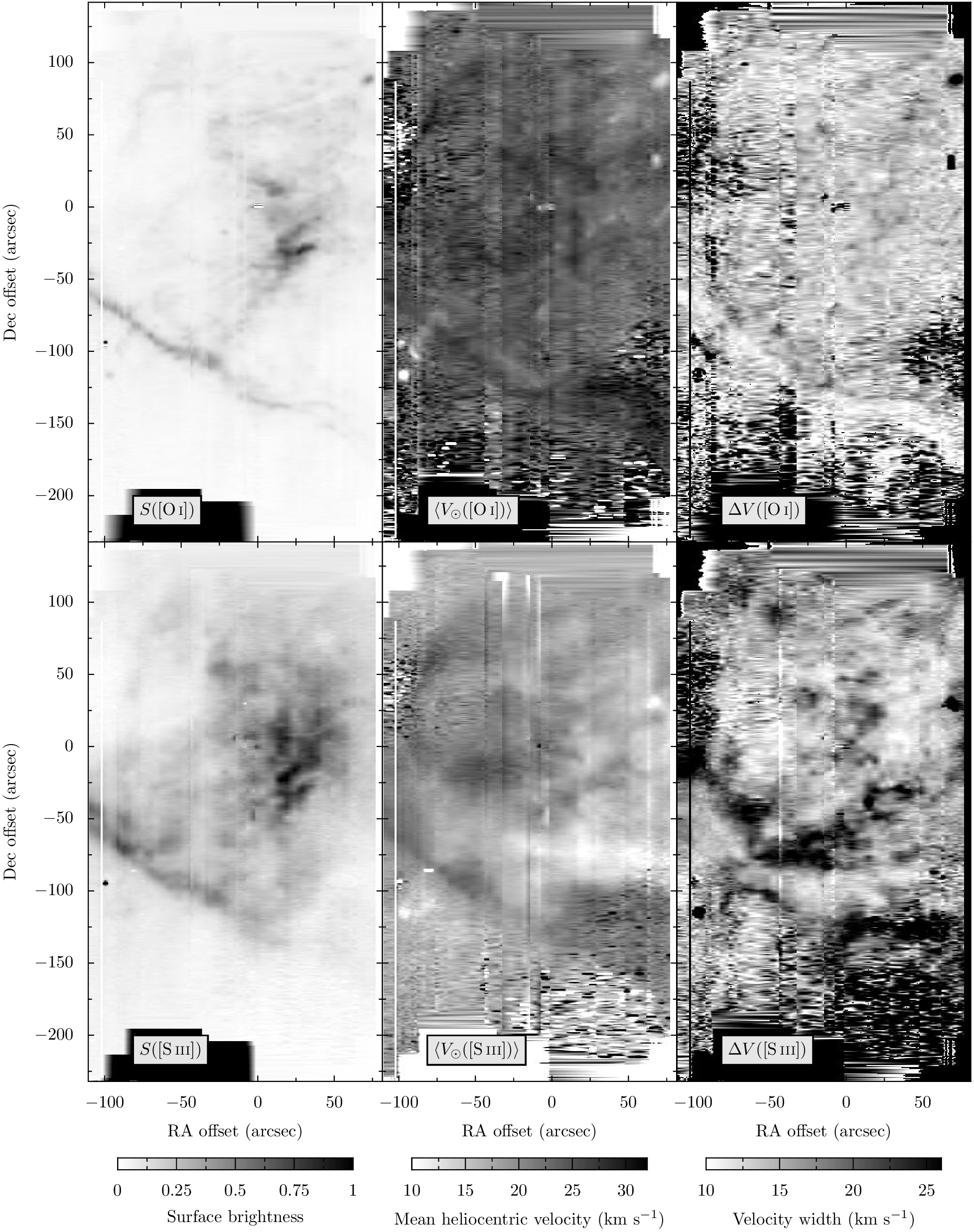}
  \caption{Emission line maps for [\ion{O}{1}]~6300\,\AA{} (top row)
    and [\ion{S}{3}]~6312\,\AA{} (bottom row) constructed from
    longslit echelle spectroscopy. Left panels show the surface
    brightness. Central panels show the mean heliocentric line
    velocity. Right panels show the FWHM line width corrected for
    thermal and instrumental broadening.}
  \label{fig:orionvel}
\end{figure*}

In order to compare with the gas-kinematic behavior of the numerical
models, we here present kinematic maps of the optical emission lines
[\ion{O}{1}]~6300\,\AA{} and [\ion{S}{3}]~6312\,\AA{}
(Figure~\ref{fig:orionvel}). These are derived from echelle
spectroscopy through a NS-oriented slit at a series of 60 different
pointings using the Mezcal spectrograph of the Observatorio
Astronómico Nacional, San Pedro Mártir, México. The observations are
described in more detail in \citet{GarciaHenney2005}. We use these
data in preference to the [\ion{N}{2}], H$\alpha$, and [\ion{O}{3}]
data of \citet{2004Doi-kinematics} because the latter contain no
information on the east-west variation of the line velocities since
they assume that the peak velocity of the total profile summed over
each NS slit is the same. With our data, on the other hand, we have
taken a pair of EW-oriented slit spectra, which allow us to ``tie
together'' the wavelength calibrations of the individual NS slits. In
addition, the existence in our spectra of the night-sky component to
the [\ion{O}{1}] line (at rest in the geocentric frame) allows us to
make a very accurate wavelength calibration for that line. The images
in Figure~\ref{fig:orionvel} show the spatial variation in line
surface brightness (which has not been corrected for foreground
extinction), the intensity-weighted mean line velocity, and the
non-thermal component of the full-width half-maximum (FWHM) of the
line, which was calculated by the subtraction in quadrature from the
observed width of the instrumental width ($9\,\kms$) and the thermal
width ($5.35\,\kms$ for [\ion{O}{1}], $3.78\,\kms$ for [\ion{S}{3}]).
All quantities were calculated for a spectral window that extends
between heliocentric velocities of $-12$ to $+40\,\kms$ to minimize
the contamination from high-velocity emission from
shocks.\footnote{Although this is successful at eliminating the highly
  blue-shifted emission associated with Herbig-Haro jets, there is
  also intermediate-velocity emission associated with the bowshocks of
  these objects that does enter our velocity window and skews the
  velocity maps in localized patches.}

The [\ion{O}{1}] line is an example of a neutral collisional line, as
discussed in \S~\ref{sec:optic-line-emiss}. As such, it
originates in a thin zone close to the ionization front and shows a
lot of fine-scale structure in its surface brightness.  The mean
velocity of [\ion{O}{1}] is only blueshifted by a few $\kms$ with
respect to the molecular gas, which emits at $25$--$28\,\kms$
\citep{1994ApJ...430..256O,2001ApJ...557..240W}. There is a general
trend for the [\ion{O}{1}] emission to become more blue-shifted
towards the west of the Trapezium, which is also seen in CO emission
from the molecular gas \citep{2001ApJ...557..240W}. Some of the bright linear
filaments seen in the surface brightness map (such as the Bright Bar
in the SW) correspond to filaments of increased redshift in the mean
velocity map. Other linear structures in the mean velocity have no
visible counterpart in the surface brightness
\citep{GarciaHenney2005}. The most redshifted [\ion{O}{1}] comes from
diffuse emission just above the SE end of the Bright Bar, while the
most blueshifted is due to the residual effect of high-velocity
shocked emission from HH\,201, 202, 203, and 204. The [\ion{O}{1}]
line width shows little large scale structure, with the notable
exception of a marked decrease in the line width along the Bright Bar
filament. Localized spots of increased line width correspond to the HH
objects discussed above.

For the ionization parameter and EUV spectrum relevant to Orion, the
[\ion{S}{3}] emission traces the fully ionized gas ($x > 0.9$), except
for the zone closest to the ionizing star, which is predominantly
S$^{3+}$. It comes from a thicker layer and so shows smoother
variations in surface brightness than does [\ion{O}{1}], being similar
in appearance to the emission measure map of Figure~\ref{fig:orion}.
The mean line velocity is blue-shifted several $\kms$ with respect to
[\ion{O}{1}] and shows a more pronounced gradient from east to west.
On the other hand, the large-scale features of the mean velocity map
are rather similar to [\ion{O}{1}], with redder emission being
associated with the emission from the Bright Bar and other linear
features. One feature seen in [\ion{S}{3}] but not in [\ion{O}{1}] is
an EW-oriented spur of blue-shifted emission to the north of the
Bright Bar \citep[the \ion{O}{3} counterpart of this feature is
discussed in][]{2004Doi-kinematics}. The average [\ion{S}{3}]
linewidth is not much different from that of [\ion{O}{1}] but
[\ion{S}{3}] shows much more variation across the nebula. The highest
widths ($15$--$20\,\kms$) are seen near the Trapezium, at the east end
of the blue spur, and in the red-shifted emission at the faint SW end
of the Bright Bar. The lowest widths ($\simeq 12\,\kms$) are seen
along the length of the Bright Bar and in a ring of radius 30--$60''$
around the Trapezium.

\section{Application of the models to Orion}
\label{sec:comp-with-simul}

\begin{figure}\centering
  \includegraphics{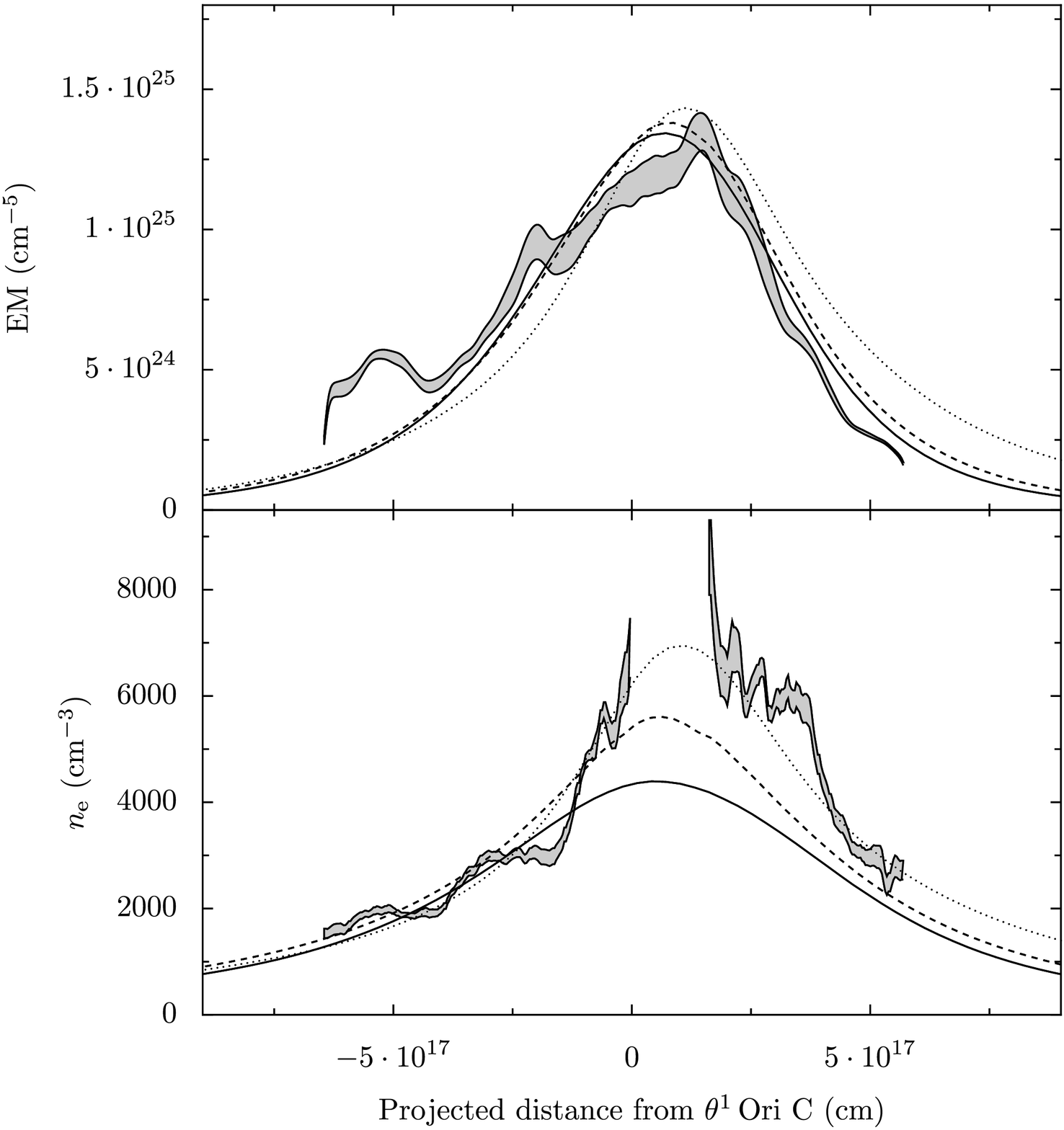}
  \caption{Observed East-West profiles of emission measure (top
    panels) and electron density (bottom panels) in the Orion nebula
    (gray line), compared with predictions of the model simulations
    for an inclination angle of $15\arcdeg$: Model~A (solid line), B
    (dashed line), and C (dotted line).}
  \label{fig:model-fits}
\end{figure}

In order to compare our numerical simulations with the observational
data of the previous section, we first extract the observational data
in an EW strip, centered on the ionizing star \thC\@. The strip is
60\arcsec{} wide in the NS direction and the average emission measure
(calculated from the H$\alpha$ surface brightness) and electron
density (calculated from the [\ion{S}{2}] line ratios) are shown in
Figure~\ref{fig:model-fits} as a function of projected distance from
the ionizing star (assuming a distance of 430\,pc,
\citet*{2001ARA&A..39...99O}). The data are represented by a gray band
that indicates $\pm3\sigma$ variation of individual pixels from the
mean. The derived electron density is very uncertain close to \thC{}
on the West side since the observed line ratio is close to the
high-density limit. We have therefore omitted this data from the
graph.

We use the three models A, B, and C at a common age of
$78,000\,\years$ as representative of our numerical results and
calculate the simulated observed profiles of emission measure and
electron density for a viewing angle of $15\arcdeg$ with respect to
the line of sight, following a procedure as close as possible to that
employed in calculating the observed maps above. For the simulated
emission measure, we first calculate simulated H$\alpha$ maps,
restricted to the line-of-sight velocity range $-4$ to $-23\,\kms$,
then convert this to emission measure by assuming a fixed H$\alpha$
emissivity corresponding to $8900\,\Kelvin$. For the electron density,
we calculate simulated maps for the $6716\,\Angstrom$ and
$6731\,\Angstrom$ [\ion{S}{2}] doublet components using the methods of
\S~\ref{sec:optic-line-emiss} for the same line-of-sight velocity
range as for H$\alpha$. We then use the ratio of these two maps to
derive $n_\mathrm{e}$ in exactly the same way as for the observations.\footnote{%
Since the ionization fraction of S$^+$ does not exactly follow that of
either neutral or ionized Hydrogen, we have used a simple fit to the
dependence of of the S$^+$ on $x$ that we derive from static
photoionization equilibrium models calculated using the Cloudy plasma
code \citep{2000RMxAC...9..153F}:
\[
\frac{n(\mathrm{S^+})}{n(\mathrm{S})} = \frac{a (1-x)^b
}{1+(a-1)(1-x)^c} , 
\]
with $a = 6.16$, $b = 0.730$, $c = 0.629$.}

For each model, we rescale the lengths and densities to the Orion
values as follows: first, we determine the star-ionization front
distance, $z_\mathrm{h,0}$, by comparing the FWHM of the width of the
EM in model units with the observed value of $6.12\times
10^{17}\,\cm$. Then, we determine the electron density scaling by
fitting by eye to the broad peak of the observed EM profile.

The results are shown in the upper panel of
Figure~\ref{fig:model-fits} in which it can be seen that all three
models can provide a reasonable fit to the emission measure profile
for a viewing angle of $15\arcdeg$. Models~A and~B fit the large-scale
skewness of the profile somewhat better than Model~C, which
overpredicts the EM on the West side and underpredicts it on the East
side. None of the models can reproduce the fine-scale structure seen
in the observed profile. In general, the agreement is better on the
West side of the profile, which is smoother than the East side.  Model
profiles for a viewing angle of $30\arcdeg$ (not shown) badly fail to
fit the observed profile since their peaks are shifted too far to the
West, while those for a viewing angle of $0\arcdeg$ (not shown) do not
fit as well as the $15\arcdeg$ profiles since they are too
symmetrical.

The parameters for the model fits in the $15\arcdeg$ case are listed
in Table~\ref{tab:fits}. These are the required effective ionizing
photon luminosity of the central star, the distances along the
symmetry axis from the star to the heating front and to the peak in
the ionized gas density (see \S~\ref{sec:numer-simul}), and the value
of the ionized density peak. Since dust grains in the ionized nebula
will absorb a fraction, $f_\mathrm{d} \sim 0.5$, of the ionizing
photons emitted by the star \citep{2004ApJ...608..282A}, and this effect
is not included in the simulations, the derived quantity is
$(1-f_\mathrm{d})Q_\mathrm{H}$.

\begin{table}
  \caption{Model fit parameters for $15\arcdeg$ viewing
    angle}\smallskip 
  \label{tab:fits}
  \setlength\tabcolsep{2.5\tabcolsep}
  \begin{tabular}{lcccc}\hline
      & $(1-f_\mathrm{d}) Q_\mathrm{H} / 10^{49}\,\mathrm{s}^{-1}$ 
      & $ z_\mathrm{h,0} / 10^{17}\,\cm$ 
      & $ z_\mathrm{p,0} / 10^{17}\,\cm$ 
      & $ n_\mathrm{p} / 10^{4}\,\pcc$\\
      Model
      & (1) & (2) & (3) & (4) \\
    \hline
    A &  0.299 & 4.42 & 3.80 & 0.707 \\
    B &  0.363 & 3.88 & 3.55 & 1.108\\
    C &  0.919 & 4.93 & 4.65 & 1.411\\
    \hline
  \end{tabular}\smallskip\footnotesize
\\
(1) Effective stellar ionizing luminosity. 
(2) Distance from star to heating front.
(3) Distance from star to density peak. 
(4) Peak ionized density. 
\end{table}

The results for the simulated observed electron density are shown in
the lower panel of Figure~\ref{fig:model-fits} where it can be seen
that Model~C does the best job of fitting the observations. The other
two models significantly underpredict the density in the central and
Western portions of the nebula. None of the models reproduce the very
high (but poorly determined) densities that seem to be indicated to
exist close to \thC{}. In \S~\ref{sec:consistency} below, we discuss
how the the geometry of Model~C can be reconciled with the fact that
the nebula appears concave on the largest scales. 

\begin{figure}\centering
  \includegraphics{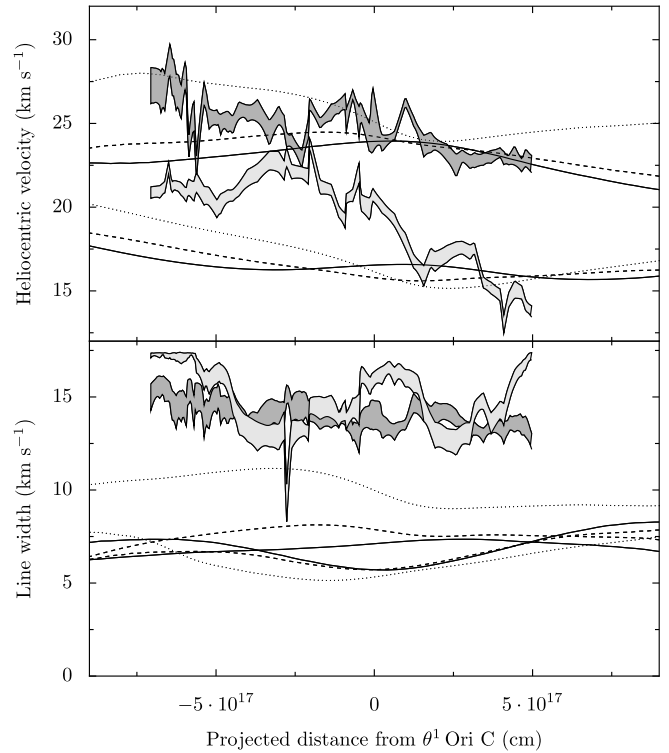}
  \caption{Observed East-West profiles of [\ion{O}{1}] (mid gray band)
    and [\ion{S}{3}] (light gray band) mean velocity (top panels) and
    FWHM (bottom panels) in the Orion nebula, compared with
    predictions of the model simulations: Model~A (solid line), B
    (dashed line), and C (dotted line).}
  \label{fig:model-vfits}
\end{figure}
We have also calculated simulated maps of mean velocity and line width
of the [\ion{O}{1}] and [\ion{S}{3}] lines in order to compare with
the observed kinematics presented in \S~\ref{sec:gas-kinematics}.
The results are shown in Figure~\ref{fig:model-vfits} for the same EW
strip as was used for the emission measure and density comparisons.
The upper panel shows the observed mean heliocentric velocities of the
two lines as gray bands indicating $\pm 3\sigma$ variation of
individual pixels at each RA\@. The lines show the model predictions
for an inclination angle of $15\arcdeg$ and assuming that the
molecular gas has a constant heliocentric velocity of $+28\,\kms$. The
lower panel shows the same but for the non-thermal FWHM of the lines. 

The models reproduce the magnitude and trend of the [\ion{O}{1}]
velocity as well as the qualitative fact that the [\ion{S}{3}] is more
blue-shifted than [\ion{O}{1}]. Although they agree with the
[\ion{S}{3}] mean velocity on the western side of the nebula, they
somewhat overpredict the blueshift of the [\ion{S}{3}] line in the
center of the nebula. Overall, Model~C is a closer fit than the other
models.

With regard to the line widths, the agreement is not very good.  The
observed [\ion{S}{3}] widths are $2$--$5\,\kms$ larger than the best
model predictions (from Model~C) and the [\ion{O}{1}] widths are
$7$--$9\,\kms$ larger than any of the model predictions.

\section{Discussion}
\label{sec:discussion}

\subsection{The Orion Nebula}
\label{sec:orion-nebula}

In \S~\ref{sec:comp-with-simul} the model predictions for the
profiles of emission measure, electron density, and line kinematic
properties are compared with observations of the Orion nebula. The EM
profile is used to calibrate the length and density scales of the
models and so its comparison with the model profiles does not provide
a strong test. It is worth noting, however, that Models~A and B can
reproduce the large-scale skewness of the EM profile for a viewing
angle of $15\arcdeg$, whereas Model~C cannot. For Model~C to produce
the observed skewness requires a larger viewing angle, which shifts
the predicted EM peak much farther to the West of \thC{} than is
observed. A further discrepancy with all the models is that the
observed profile shows much fine-scale structure that is entirely
absent in the simulations. 

Since the observed $n_\mathrm{e}$ profile was not used in determining
the model parameters, it provides a much more stringent test for the
models. Only Model~C predicts a sufficently high density in the
central and Western regions of the nebula. Again, the observations
show structure on scales $\le 10^{17}\,\cm$, which are not present in
the models. 

\subsubsection{Required ionizing photon luminosity} 

The derived parameters from the model fits (Table~\ref{tab:fits}) show
striking differences between the three models. In particular, the
required effective ionizing photon luminosity is $Q\Sub{H}
(1-f_\mathrm{d}) = 9.2\times 10^{48}\,\persecond$ for Model~C (where
$f_\mathrm{d}$ is the fraction of ionizing photons absorbed by dust),
which is roughly three times higher than the equivalent values for
Models~A and B\@. This is partly because the ionized density is more
concentrated at the ionization front in Model~C and thus requires a
higher ionizing luminosity to sustain the same emission measure as
compared to a situation in which the ionized gas is more evenly
distributed between the star and the ionization front, as is the case
for Models~A and~B\@. Another contributing factor is that the
convexity of the ionization front of Model~C (see
Figure~\ref{fig:snapshot}) means that the nebula is density-bounded
for a large fraction of the solid angle as seen from the central star,
whereas Models~A and B are more concave and thus more ionization
bounded. It has long been known that neutral gas in the so-called
``veil'' overlays the Orion nebula
\citep{1989A&A...224..209V,2004ApJ...609..247A}, so that the nebula
must be ionization bounded on a scale of a few parsecs, thus
contributing a low surface-brightness halo to the ionized emission.
However, this would make a negligible contribution to the brightness
of the inner region studied here. Indeed, \citet{1993A&AS...98..137F} find the
total effective ionizing luminosity driving the nebula to be $9\times
10^{48}\,\persecond$ from single-dish radio observations, which are
sensitive to emission on a scale of parsecs that is not detected by
interferometers. This is fully consistent with Model~C but not with
the other two models. Another independent estimate of the stellar
ionizing luminosity comes from studying the surface brightness of the
cusps of the Orion proplyds
\citep{1998AJ....116..322H,1999AJ....118.2350H}, which gives a value
of $(1-f'\Sub{d}) Q\Sub{H} = (1.0\pm0.2) \times 10^{49}\,\persecond$
for proplyd 167-317 (LV\,2) that has the best-determined parameters
\citep{2002ApJ...566..315H}. The fraction of ionizing photons absorbed
by dust in the proplyd flow, $f'\Sub{d}$, should be similar that for
the nebula as a whole, $f\Sub{d}$, since in both cases the dust
optical depth through the ionized gas is of order unity, so this
result is also consistent only with our Model~C.

\subsubsection{Emission line kinematics and broadening}
\label{sec:emiss-line-kinem}

The comparison between the predicted and observed line kinematics is
only partially successful. On the one hand, the trend of increasing
blueshift with ionization is well reproduced by the models.  This is a
vast improvement on the plane-parallel weak-D models presented in
\citet{2005ApJ...621..328H}, which were unable to produce a difference of
more than $1\,\kms$ between the neutral and fully ionized lines. Our
photoevaporation models predict a difference of about $5\,\kms$
between [\ion{O}{1}] and [\ion{S}{3}], which is very similar to what
is observed. The observations show a shift in the mean velocity from E
to W of $\simeq 3\,\kms$ for [\ion{O}{1}] and $\simeq 6\,\kms$ for
[\ion{S}{3}], which are both approximately reproduced by the models.
In the case of [\ion{O}{1}], some part of this variation can be
accounted for by shifts in the velocity of the molecular gas behind
the ionization front \citep{2001ApJ...557..240W} and furthermore the
westernmost portion of the [\ion{S}{3}] is affected by
intermediate-velocity gas associated with the HH\,202 jet and
bowshock. Taking these into account, Model~C can be seen to fit the
observations quite well, apart from in the region around the Trapezium
where the observed [\ion{S}{3}] velocity is about $4\,\kms$ redder
than the model prediction. This may be evidence for the interaction of
the stellar wind from \thC{} with the photoevaporation flow, as is
discussed further below.

On the other hand, the discrepancy seen between the predicted and
observed line widths is very striking, especially for [\ion{O}{1}],
for which the models all predict a FWHM of $7\,\kms$, which is only
half the observed value. For [\ion{S}{3}], the discrepancy, although
less, is still significant: the best model predicts a FWHM of
$12\,\kms$ whereas the observed values range from $13$ to $17\,\kms$.
Since the widths of independent broadening agents should add
approximately in quadrature, we thus require an extra broadening of
$12\,\kms$ for [\ion{O}{1}] and $9\,\kms$ for [\ion{S}{3}]. The line
widths in the models are mainly determined by the physical velocity
gradients, $dv/dz$, in the photoevaporation flow and are of magnitude
$\sim \left\vert dv/dz \right\vert \delta z$, where $ \delta z$ is the
thickness of the emitting layer. In principle, an additional
broadening agent could be variation along the line of sight of the
\emph{direction} of the flow velocity vector (with magnitude $\sim
\left\vert v \right\vert \delta\theta$) but this effect is very small
in our models since the radius of curvature of the ionization front is
of order the size of the nebula, so $\delta\theta$ is small. This
broadening would be much more important if the ionization front had
many small-scale irregularities, which does seem to be indicated by
the large amount of structure seen in the surface brightness maps
\citep{1991ApJ...369L..75H,1991PASP..103..824O,2003AJ....125.2590O}.
An example of where this effect has been shown to be important is in
the flows from the Orion proplyds \citep{1999AJ....118.2350H}. These
are poorly resolved spatially with ground-based spectroscopy, so the
full range of angles is present and one predicts line widths of order
$2\left\vert v \right\vert$, with photoevaporation models successfully
reproducing the observed line profiles for fully ionized species.
However, although this small-scale divergence broadening might
plausibly explain the [\ion{S}{3}] linewidth discrepancy, it is hard
to see how it can work for [\ion{O}{1}] since the magnitude of the
velocity in the [\ion{O}{1}]-emitting zone is only $2$--$3\,\kms$,
which, even when doubled, is much smaller than the required broadening
of $12\,\kms$.\footnote{The [\ion{O}{1}] linewidths from the proplyds
  are also inexplicable in terms of kinematic broadening since they
  are very similar to the widths in the surrounding nebula
  \citep{1999AJ....118.2350H,2001ApJ...556..203O}.}  Similar arguments
apply to the contribution to the observed widths of back-scattered
emission from dust behind the emitting layer
\citep{1998ApJ...503..760H}, which again may be important for
blueshifted lines such as [\ion{S}{3}] but will be unimportant for
[\ion{O}{1}].

Previous authors have reached similar conclusions about the need for
an extra broadening agent in the nebula from a purely empirical
viewpoint \citep[see][ \S~3.4.2--4 and references
therein]{2001ARA&A..39...99O}. More recently,
\citet{2003AJ....125.2590O} have found evidence that the extra
broadening may be greater in recombination lines than in collisional
lines and thus may be related to the equally puzzling problem of
temperature variations in the nebula.

Another proposed solution is that the broadening is caused by Alfvén
waves, as is believed to be the case in giant molecular clouds
\citep{1988ApJ...326L..27M}. In order to give an Alfvén velocity
($v\Sub{a} = \left(B^2/4\pi\rho\right)^{1/2}$) of $9\,\kms$, which is
necessary to explain the observed [\ion{O}{1}] broadening, one
requires a magnetic field of order $B = 0.5\,\mathrm{mG}$ in the
region where the line forms (which has an ionization fraction of order
20\% and a total gas density roughly 3 times the peak ionized
density). This is marginally inconsistent with the upper limit on $B$
derived from the Faraday rotation measure \citep{1998ApJ...502L..75R}
and, furthermore, would predict very high Alfvén speeds either in the
dense molecular gas or in the lower density, highly ionized gas.  To
see this, consider two possibilities for how the magnetic field
responds to compression/rarefaction: first, suppose that $B$ is
constant, such as might be the case for a highly ordered field with
orientation parallel to the flow direction (perpendicular to the
ionization front). In this case, $v\Sub{A} \sim \rho^{-1/2}$ and so a
reasonable value is obtained in the dense molecular gas ($v\Sub{A}
\simeq 3\,\kms$ at $10^5\,\pcc$) but a very high value is obtained in
the more highly ionized, lower density regions of the flow ($v\Sub{A}
\simeq 30\,\kms$ at $10^3\,\pcc$), which is ruled out by observations
of lines such as [\ion{O}{3}].  Second, suppose that $B$ is tangled or
chaotic and can be considered as an isotropic pressure with effective
adiabatic index $\gamma\Sub{m} = 4/3$ \citep[see][
Appendix~C]{2005ApJ...621..328H}. In this case, $v\Sub{A} \sim
\rho^{(\gamma\Sub{m}-1)/2} \sim \rho^{1/6}$, so that the Alfvén speed
would have to exceed $10\,\kms$ in the molecular gas, which greatly
exceeds the linewidths of the molecular lines. Although we cannot
definitely rule out the role of Alfvén waves, these considerations
suggest that they do not provide a natural explanation for the line
broadening.

\subsubsection{Consistency with the large-scale nebular geometry}
\label{sec:consistency}
\begin{figure*}
  \makebox[0.5\linewidth][l]{(\textit{a})}%
  \makebox[0.5\linewidth][l]{(\textit{b})}\\
  \setkeys{Gin}{width=0.5\textwidth}
  \includegraphics{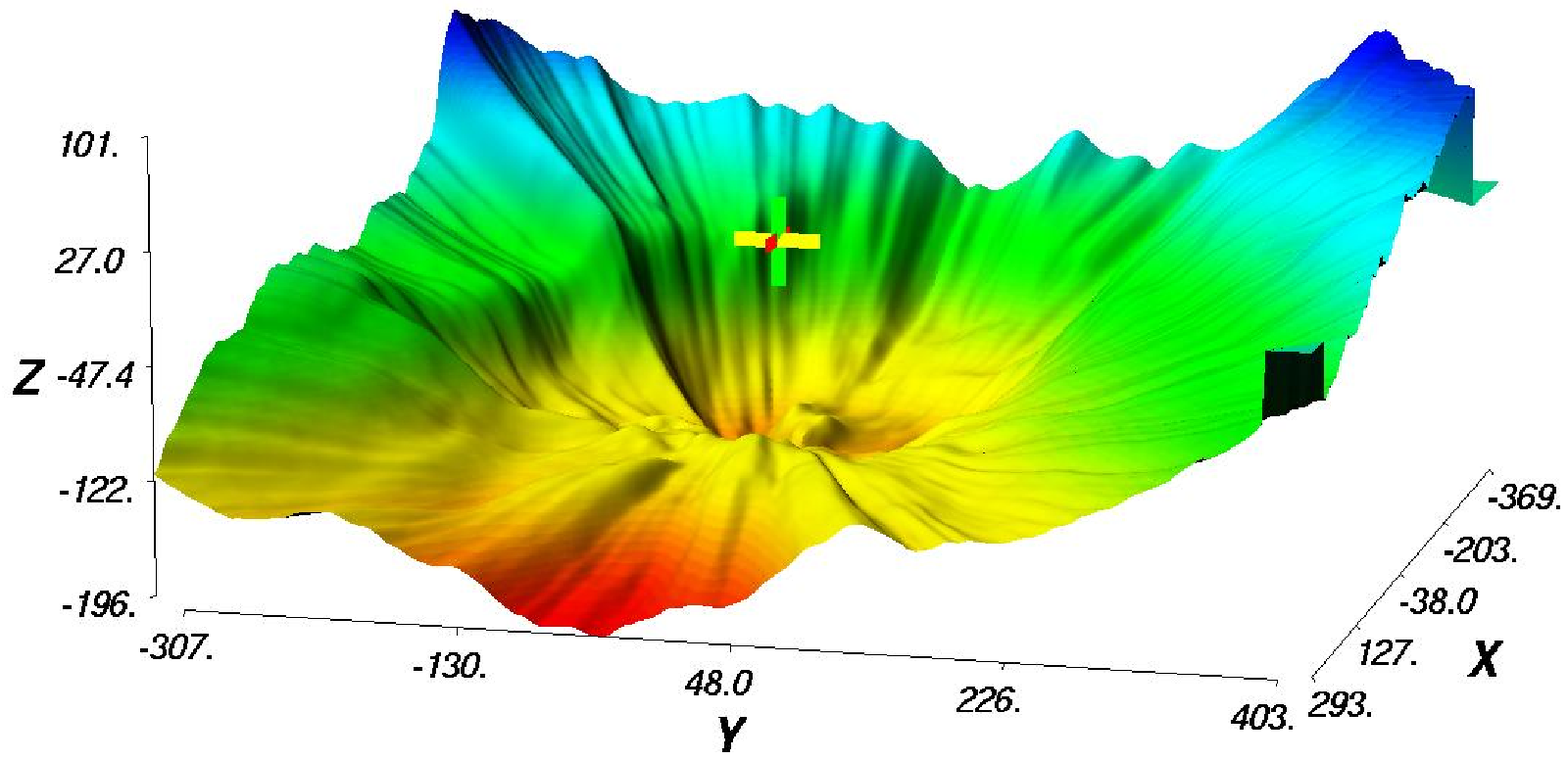}\includegraphics{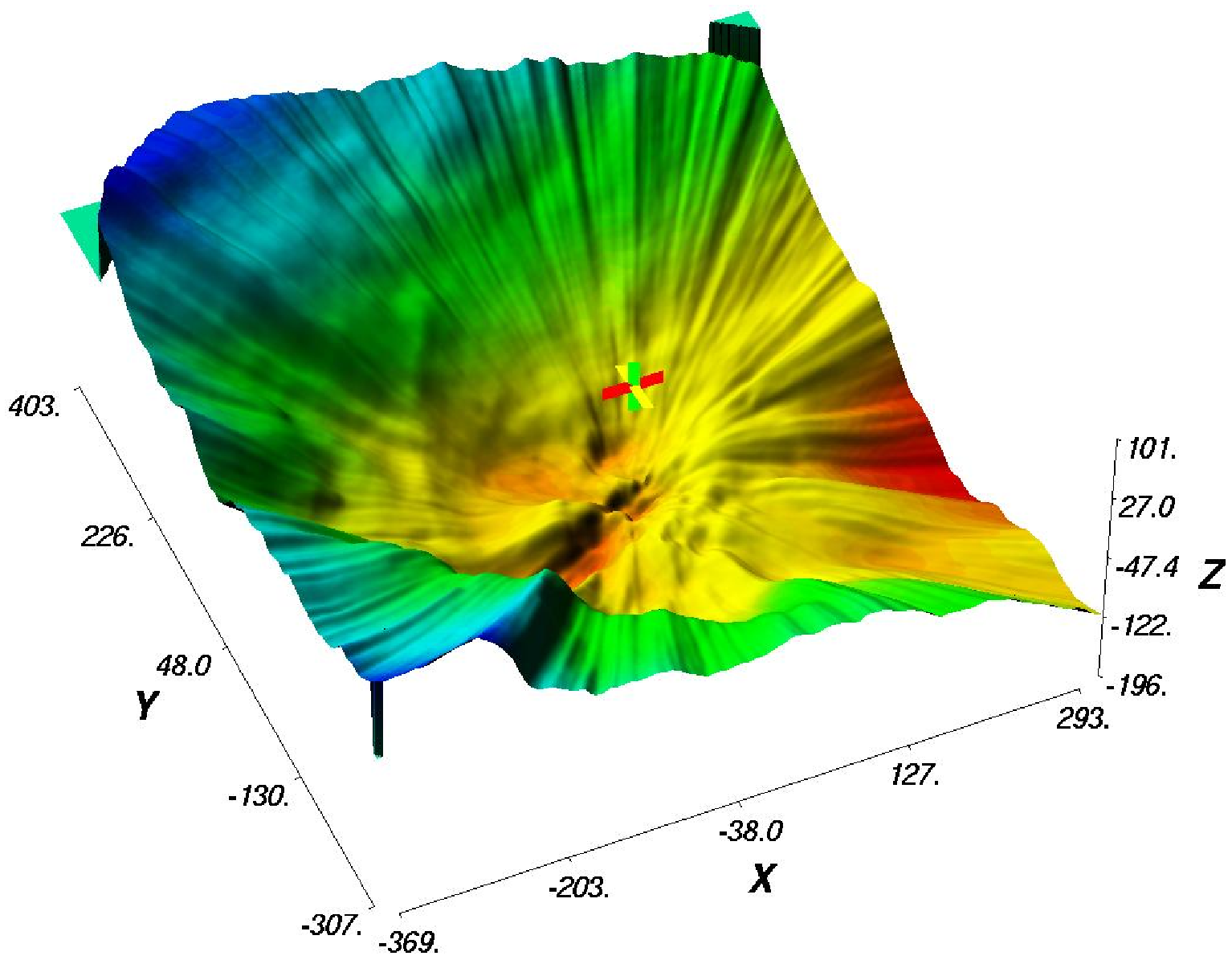}
  \caption{Three-dimensional shape of the principal ionization front
    in Orion, derived from radio free-free surface brightness
    \citep{Henney2005-3D}.  (\textit{a})~View over the molecular ridge
    from the WNW direction.  (\textit{b})~View over the Bright bar
    from the SSE direction.  The $x$, $y$, and $z$ axes correspond to
    displacements along the EW, NS, and line-of-sight directions,
    respectively, with numerical values marked in arcseconds ($1'' =
    6.45 \times 10^{15}\,\cm$). The ionizing star is at the origin and
    is marked by a cross, while the observer is at $(x,y,z) =
    (0,0,+\infty)$.  Shading of the surface indicates the $z$ height
    from low (red) to high (blue).  This figure is available in color
    in the online edition.}
  \label{fig:3d}
\end{figure*}
Comparison of our models with the observed electron densities and line
velocities in the nebula strongly favor Model~C over the other two
models, so it is worth considering whether the shape of the ionization
front in this model is consistent with independent determinations. In
Model~C, the ionization front has a positive curvature, that is, it is
convex as seen from the ionizing star. However, three-dimensional
reconstruction of the shape of the ionization front paints a somewhat
more complex picture \citep{1995ApJ...438..784W, Henney2005-3D}, as is
shown in Figure~\ref{fig:3d}. In the brightest part of the nebula,
just to the W of the ionizing star, the ionization front is indeed
locally convex in the EW direction but is roughly flat in the NS
direction.\footnote{The front seems to wrap around the dense molecular
  filament that is seen in molecular line and dust continuum
  observations \citep{1999ApJ...510L..49J}.} Also, the axis between the
star and the closest point on the front is inclined at about
$15\arcdeg$ to the line of sight, consistent with what we found for
our models. On a larger scale, however, the front becomes concave in
all directions apart from the south-west. The implications for our
models are twofold: first, on a medium scale ($< 1\arcmin \simeq 4
\times 10^{17}\,\cm$) the flow should not be strictly cylindrically
symmetric but instead should resemble Model~C in its EW cross-section
but more resemble Model~B in its NS cross-section. Since we compared
our models against an EW strip of the nebula, it is reasonable that
Model~C produced better fits. Secondly, on a larger scale the flow
should come to more resemble Model~A, but with a symmetry axis that is
inclined towards the south-west, rather than towards the east and
possible at a greater angle to the line-of-sight.\footnote{This is
  also consistent with the parsec-scale optical appearance of the
  nebula, in which the flow appears to be towards the south-east.} How
the flow will adjust between the differing curvature and symmetry axes
is unclear, but it may involve internal shocks. Further complications
will arise due to the interaction of the global flow with local flows
from other convex features in the ionization front. The largest and
most prominent of these is the Bright Bar but there are many other
examples \citep{2000AJ....120..382O}, some of which are apparent in
Figure~\ref{fig:3d}.

\subsection{Shortcomings of our models}
\label{sec:things-we-left-out}

Although our photoevaporation flow models have had some measure of
success in accounting for the structure and kinematics of the Orion
nebula on medium scales, there are many physical processes that we
have neglected and which would need to be included in a complete
model. 

We have used a very simplified prescription for the radiative transfer
in our models, in which we consider only one frequency for the
ionizing radiation. Although we do include the hardening of the
radiation field, thus accounting for the important reduction in the
photoionization cross section deep in the ionization front, it would
be more satisfactory to use a multi-frequency approach. The
consideration of higher energy EUV radiation would allow us to
calculate the helium ionization structure, while the inclusion of
non-ionizing FUV radiation would permit a more realistic treatment of
the heating and cooling in the neutral gas. Furthermore, the effects
of dust absorption on the transfer of ionizing radiation are important
and should be included.

One failure of our model has been an inability to acount for the
reduced blueshift of high-ionization lines that is seen within a few
times $10^{17}\,\cm$ of the Trapezium. It is possible that the stellar
wind from \thC{} or radiation pressure acting on dust grains may be
important in this inner region. Both are effects that need to be
incorporated in a future model. 

There is much small scale structure present in the Orion ionization
front \citep{1991ApJ...369L..75H}, right down to scales $\sim
10^{15}\,\cm$,\footnote{It is unclear whether this structure reflects
  pre-existing structure in the molecular cloud or whether it is due
  to instabilities associated with the ionization front
  \citep{1996ApJ...469..171G, 1999MNRAS.310..789W,
    2002MNRAS.331..693W}.} or smaller still if one includes the
proplyds.  Our models can be taken as reflecting the mean flow
properties after all this small scale structure has been smeared out,
but it is not obvious that such an approach is strictly valid. The
very highest electron densities found close to the Trapezium are not
reproduced by our models and seem to require that the emission there
is dominated by much smaller-scale flows than we have considered. The
effects of the mutual interaction of such flows is still in need of
further research \citep[but see][]{2003RMxAC..15..175H}, which also
needs to address the influence of bar-like features and the change in
the orientation of the symmetry axis when one passes from
$10^{18}\,\cm$ to $10^{19}\,\cm$ scales.

Lastly, but perhaps most importantly, our models completely neglect
the role of the magnetic field. In the cold molecular gas, the
pressure associated with magnetic field is expected to dominate the
gas pressure, and so should have an important dynamic effect, but it
is unclear if this is still the case in the ionized gas. For most
plausible assumptions about the magnetic field geometry, the magnetic
pressure is found to be a small fraction of the thermal pressure in
the ionized gas \citep{2005ApJ...621..328H}. If the unexplained
broadening of the [\ion{O}{1}] line found in
\S~\ref{sec:gas-kinematics} is ascribed to Alfvén turbulence,
then this fraction may be as high as 25\% at the ionization front.
However, even if the magnetic field is dynamically unimportant in the
ionized gas, it can still have an important effect on the jump
conditions at the ionization front \citep{1998A&A...331.1099R,
  2000MNRAS.314..315W, 2001MNRAS.325..293W}.

\section{Conclusions}
\label{sec:conclusions}

We have carried out two-dimensional, cylindrically symmetric,
hydrodynamical simulations of the photoevaporation of a cloud with
large-scale density gradients, which gives rise to a transonic,
ionized, photoevaporation flow with the following general properties:
\begin{enumerate}
\item After an initial transient phase, with duration $\sim
  10^4\,\years$ for typical compact \hii{} region sizes
  (\S~\ref{sec:early-model-evolution}), the flow enters a long-lived
  quasi-stationary phase in which the ionized flow is approximately
  steady in the frame of reference of the ionization/heating front
  (\S~\ref{sec:quasi-steady}).
\item During the quasi-stationary phase (\S\S~\ref{sec:quasi-steady},
  \ref{sec:secul-evol-flow}), the flow structure is determined
  entirely by two parameters: the distance of the ionizing star from
  the front, $z_0$, and the curvature of the front, $\kappa$.
  Depending on the curvature, the flows are more ``champagne-like''
  ($\kappa < 0$) or ``globule-like'' ($\kappa > 0$).
\item The curvature of the front (\S~\ref{sec:secul-evol-flow})
  depends on the lateral density distribution in the neutral gas and
  on the evolutionary stage of the flow. Positively curved fronts are
  only obtained for lateral density distributions that are
  asymptotically steeper than $r^{-2}$. We found that initially flat
  or concave fronts tend to evolve with time towards more negative
  values of $\kappa z_0$ (increasing concavity), whereas initially
  convex fronts evolve towards more positive values of $\kappa z_0$
  (increasing convexity).
\item The flat and convex ionization fronts in the simulations are
  found to be \hbox{D-critical}, whereas concave fronts are found to be
  weak-D (Appendix~\ref{sec:class-ioniz-front}). There is still a
  sonic point in the flows from concave fronts but it occurs in fully
  ionized gas, rather than in the front itself.
\item The relationship between gas ionization and kinematics shows
  little variation as a function of the front curvature
  (\S~\ref{sec:optic-line-emiss}). In all cases, a gradient of $\simeq
  10\,\kms$ between the neutral and ionized gas is obtained. There is,
  however, a tendency for models with more positive curvature to show
  larger mean velocities and line widths for ionized species.
\end{enumerate}

We have compared our models with both new and existing observations of
the Orion nebula (\S\S~\ref{sec:observational-data},
\ref{sec:comp-with-simul}), with the following results:

\begin{enumerate}
\item Only a convex model can simultaneously reproduce the observed
  distributions of emission measure and electron density in the
  nebula. A good fit is obtained for $\kappa z_0 \simeq 0.5$, implying
  that the radius of curvature of the front is roughly twice its
  distance from the ionizing star. Flat and concave models predict an
  electron density in the west of the nebula that is much lower than
  is observed. They do, however, fit the observed emission measure
  distribution slightly better than a convex model.
\item Only a convex model is consistent with independent estimates of
  the ionizing luminosity, $Q\Sub{H}$, from \thC{}. Flat and concave
  models trap a much higher fraction of the stellar ionizing photons
  in the central region of the nebula and predict a $Q\Sub{H}$ that is
  three times too small.
\item All the models can broadly reproduce the observed mean
  velocities of optical [\ion{O}{1}] and [\ion{S}{3}] emission lines
  but a convex model fits the data better.
\item None of the models can reproduce the observed widths of the
  optical emission lines, although convex models come close to doing
  so for the high-ionization lines. The largest discrepancy is found
  for the neutral [\ion{O}{1}] line, which requires an extra
  broadening agent with a FWHM of $12\,\kms$.
\item The best-fitting model over all has a flow axis inclined
  $15\arcdeg$ west from the line of sight and a distance between
  \thC{} and the ionization front of $4.7 \times 10^{17}\,\cm$. 
\end{enumerate}

\acknowledgments

We would like to thank Bob O'Dell for freely sharing his observational
data and for many discussions about the Orion nebula. We also thank
Robin Williams, Alex Raga, Gary Ferland, and Jeff Hester for useful
discussions, and the anonymous referee for a most thorough and helpful
report.  This work has made use of the observational facilities at San
Pedro Mártir Observatory, B.C., Mexico, and NASA's Astrophysics Data
System.  We are grateful for financial support from DGAPA-UNAM,
Mexico, through project IN115202 and through sabbatical grants to WJH
and SJA\@. MTGD thanks CONACyT, Mexico, for support through a research
studentship. We are also grateful to the University of Leeds, UK, for
hospitality during a year-long sabbatical visit, during which this
work was begun.

\bibliographystyle{astroads}
\bibliography{%
  HII-2003,HII-2002,HII-2001,HII-2000,HII-1999,HII-1998,HII-1997,HII-1996,Ferland,WillPapers,Extras,IfrontDynamics,all-odell,vandervoort,bertoldi,Kahn,Orion-CO,Pepe,TenorioTagle,Hester}

\begin{thebibliography}{83}
\expandafter\ifx\csname natexlab\endcsname\relax\def\natexlab#1{#1}\fi
\expandafter\ifx\csname href\endcsname\relax
  \def\href#1#2{}\fi
\expandafter\ifx\csname urllinklabel\endcsname\relax
  \def\urllinklabel{[LINK]}\fi
\expandafter\ifx\csname adsurllinklabel\endcsname\relax
  \def\adsurllinklabel{[ADS]}\fi

\bibitem[{{Abel} {et~al.}(2004){Abel}, {Brogan}, {Ferland}, {O'Dell}, {Shaw},
  \& {Troland}}]{2004ApJ...609..247A}
{Abel}, N.~P., {Brogan}, C.~L., {Ferland}, G.~J., {O'Dell}, C.~R., {Shaw}, G.,
  \& {Troland}, T.~H. 2004, \apj, 609, 247


\bibitem[{{Arthur} {et~al.}(2004){Arthur}, {Kurtz}, {Franco}, \& {Albarr{\'
  a}n}}]{2004ApJ...608..282A}
{Arthur}, S.~J., {Kurtz}, S.~E., {Franco}, J., \& {Albarr{\' a}n}, M.~Y. 2004,
  \apj, 608, 282


\bibitem[{{Baldwin} {et~al.}(1991){Baldwin}, {Ferland}, {Martin}, {Corbin},
  {Cota}, {Peterson}, \& {Slettebak}}]{1991ApJ...374..580B}
{Baldwin}, J.~A., {Ferland}, G.~J., {Martin}, P.~G., {Corbin}, M.~R., {Cota},
  S.~A., {Peterson}, B.~M., \& {Slettebak}, A. 1991, \apj, 374, 580
 \href{http://adsabs.harvard.edu/cgi-bin/nph-bib_query?bibcode=1991ApJ...374..%
580B&db_key=AST}{\adsurllinklabel}

\bibitem[{{Balick} {et~al.}(1974){Balick}, {Gammon}, \&
  {Hjellming}}]{1974PASP...86..616B}
{Balick}, B., {Gammon}, R.~H., \& {Hjellming}, R.~M. 1974, \pasp, 86, 616


\bibitem[{{Bally} {et~al.}(2000){Bally}, {O'Dell}, \&
  {McCaughrean}}]{2000AJ....119.2919B}
{Bally}, J., {O'Dell}, C.~R., \& {McCaughrean}, M.~J. 2000, \aj, 119, 2919


\bibitem[{{Bedijn} \& {Tenorio-Tagle}(1981)}]{1981A&A....98...85B}
{Bedijn}, P.~J. \& {Tenorio-Tagle}, G. 1981, \aap, 98, 85


\bibitem[{{Bedijn} \& {Tenorio-Tagle}(1984)}]{1984A&A...135...81B}
---. 1984, \aap, 135, 81


\bibitem[{{Bertoldi}(1989)}]{1989ApJ...346..735B}
{Bertoldi}, F. 1989, \apj, 346, 735


\bibitem[{{Bertoldi} \& {McKee}(1990)}]{1990ApJ...354..529B}
{Bertoldi}, F. \& {McKee}, C.~F. 1990, \apj, 354, 529


\bibitem[{{Bodenheimer} {et~al.}(1979){Bodenheimer}, {Tenorio-Tagle}, \&
  {Yorke}}]{1979ApJ...233...85B}
{Bodenheimer}, P., {Tenorio-Tagle}, G., \& {Yorke}, H.~W. 1979, \apj, 233, 85


\bibitem[{{Cai} \& {Pradhan}(1993)}]{1993ApJS...88..329C}
{Cai}, W. \& {Pradhan}, A.~K. 1993, \apjs, 88, 329


\bibitem[{{Cardelli} {et~al.}(1989){Cardelli}, {Clayton}, \&
  {Mathis}}]{1989ApJ...345..245C}
{Cardelli}, J.~A., {Clayton}, G.~C., \& {Mathis}, J.~S. 1989, \apj, 345, 245


\bibitem[{{Comeron}(1997)}]{1997A&A...326.1195C}
{Comeron}, F. 1997, \aap, 326, 1195
 \href{http://adsabs.harvard.edu/cgi-bin/nph-bib_query?bibcode=1997A%26A...326%
.1195C&db_key=AST}{\adsurllinklabel}

\bibitem[{{Doi} {et~al.}(2004){Doi}, {O'Dell}, \&
  {Hartigan}}]{2004Doi-kinematics}
{Doi}, T., {O'Dell}, C.~R., \& {Hartigan}, P. 2004, \aj, 0, 0


\bibitem[{{Dyson}(1968)}]{1968Ap&SS...1..388D}
{Dyson}, J.~E. 1968, \apss, 1, 388


\bibitem[{{Felli} {et~al.}(1993){Felli}, {Churchwell}, {Wilson}, \&
  {Taylor}}]{1993A&AS...98..137F}
{Felli}, M., {Churchwell}, E., {Wilson}, T.~L., \& {Taylor}, G.~B. 1993, \aaps,
  98, 137


\bibitem[{{Ferland}(2000)}]{2000RMxAC...9..153F}
{Ferland}, G.~J. 2000, in Astrophysical Plasmas: Codes, Models, and
  Observations, (Eds. S. J. Arthur, N. Brickhouse, and J. Franco) Revista
  Mexicana de Astronom{\'{\i}}a y Astrof{\'{\i}}sica (Serie de Conferencias),
  Vol.~9, 153--157
 \href{http://adsabs.harvard.edu/cgi-bin/nph-bib_query?bibcode=2000RMxAC...9..%
153F&db_key=AST}{\adsurllinklabel}

\bibitem[{{Ferland}(2001)}]{2001PASP..113...41F}
{Ferland}, G.~J. 2001, \pasp, 113, 41
 \href{http://adsabs.harvard.edu/cgi-bin/nph-bib_query?bibcode=2001PASP..113..%
.41F&db_key=AST}{\adsurllinklabel}

\bibitem[{{Franco} {et~al.}(1989){Franco}, {Tenorio-Tagle}, \&
  {Bodenheimer}}]{1989RMxAA..18...65F}
{Franco}, J., {Tenorio-Tagle}, G., \& {Bodenheimer}, P. 1989, Revista Mexicana
  de Astronomia y Astrofisica, vol.~18, 18, 65


\bibitem[{{Garc\'\i{}a D\'\i{}az} \& {Henney}(2003)}]{garcia03}
{Garc\'\i{}a D\'\i{}az}, M.~T. \& {Henney}, W.~J. 2003, in Winds, Bubbles and
  Explosions, (Eds. S. J. Arthur and W. J. Henney) Revista Mexicana de
  Astronom{\'{\i}}a y Astrof{\'{\i}}sica (Serie de Conferencias), Vol.~15,
  201--201


\bibitem[{{Garc\'\i{}a D\'\i{}az} \& {Henney}(2005)}]{GarciaHenney2005}
{Garc\'\i{}a D\'\i{}az}, M.~T. \& {Henney}, W.~J. 2005, \apj, in preparation


\bibitem[{{Garcia-Segura} \& {Franco}(1996)}]{1996ApJ...469..171G}
{Garcia-Segura}, G. \& {Franco}, J. 1996, \apj, 469, 171


\bibitem[{Godunov(1959)}]{Godunov}
Godunov, S.~K. 1959, Mat. Sbornik, 47, 271


\bibitem[{{Goldsworthy}(1961)}]{Goldsworthy-1961}
{Goldsworthy}, F.~A. 1961, Phil. Trans. A, 253, 277


\bibitem[{{Henney}(1998)}]{1998ApJ...503..760H}
{Henney}, W.~J. 1998, \apj, 503, 760
 \href{http://adsabs.harvard.edu/cgi-bin/nph-bib_query?bibcode=1998ApJ...503..%
760H&db_key=AST}{\adsurllinklabel}

\bibitem[{{Henney}(2001)}]{2001RMxAC..10...57H}
{Henney}, W.~J. 2001, in The Seventh Texas-Mexico Conference on Astrophysics:
  Flows, Blows and Glows (Eds. William H. Lee and Silvia Torres-Peimbert)
  Revista Mexicana de Astronom{\'{\i}}a y Astrof{\'{\i}}sica (Serie de
  Conferencias) Vol. 10, pp. 57-60 (2001), 57--60
 \href{http://adsabs.harvard.edu/cgi-bin/nph-bib_query?bibcode=2001RMxAC..10..%
.57H&db_key=AST}{\adsurllinklabel}

\bibitem[{{Henney}(2003)}]{2003RMxAC..15..175H}
{Henney}, W.~J. 2003, in Winds, Bubbles and Explosions, (Eds. S. J. Arthur and
  W. J. Henney) Revista Mexicana de Astronom{\'{\i}}a y Astrof{\'{\i}}sica
  (Serie de Conferencias), Vol.~15, 175--180


\bibitem[{Henney(2005)}]{Henney2005-3D}
Henney, W.~J. 2005, \apj, xxx, xxx, in preparation


\bibitem[{{Henney} \& {Arthur}(1998)}]{1998AJ....116..322H}
{Henney}, W.~J. \& {Arthur}, S.~J. 1998, \aj, 116, 322
 \href{http://adsabs.harvard.edu/cgi-bin/nph-bib_query?bibcode=1998AJ....116..%
322H&db_key=AST}{\adsurllinklabel}

\bibitem[{{Henney} {et~al.}(2005){Henney}, {Arthur}, {Williams}, \&
  {Ferland}}]{2005ApJ...621..328H}
{Henney}, W.~J., {Arthur}, S.~J., {Williams}, R.~J.~R., \& {Ferland}, G.~J.
  2005, \apj, 621, 328


\bibitem[{{Henney} \& {O'Dell}(1999)}]{1999AJ....118.2350H}
{Henney}, W.~J. \& {O'Dell}, C.~R. 1999, \aj, 118, 2350
 \href{http://adsabs.harvard.edu/cgi-bin/nph-bib_query?bibcode=1999AJ....118.2%
350H&db_key=AST}{\adsurllinklabel}

\bibitem[{{Henney} {et~al.}(2002){Henney}, {O'Dell}, {Meaburn}, {Garrington},
  \& {Lopez}}]{2002ApJ...566..315H}
{Henney}, W.~J., {O'Dell}, C.~R., {Meaburn}, J., {Garrington}, S.~T., \&
  {Lopez}, J.~A. 2002, \apj, 566, 315
 \href{http://adsabs.harvard.edu/cgi-bin/nph-bib_query?bibcode=2002ApJ...566..%
315H&db_key=AST}{\adsurllinklabel}

\bibitem[{{Hester} {et~al.}(1991){Hester}, {Gilmozzi}, {O'Dell}, {Faber},
  {Campbell}, {Code}, {Currie}, {Danielson}, {Ewald}, {Groth}, {Holtzman},
  {Kelsall}, {Lauer}, {Light}, {Lynds}, {O'Neil}, {Shaya}, \&
  {Westphal}}]{1991ApJ...369L..75H}
{Hester}, J.~J., {Gilmozzi}, R., {O'Dell}, C.~R., {Faber}, S.~M., {Campbell},
  B., {Code}, A., {Currie}, D.~G., {Danielson}, G.~E., {Ewald}, S.~P., {Groth},
  E.~J., {Holtzman}, J.~A., {Kelsall}, T., {Lauer}, T.~R., {Light}, R.~M.,
  {Lynds}, R., {O'Neil}, E.~J., {Shaya}, E.~J., \& {Westphal}, J.~A. 1991,
  \apjl, 369, L75


\bibitem[{{Hester} {et~al.}(1996){Hester}, {Scowen}, {Sankrit}, {Lauer},
  {Ajhar}, {Baum}, {Code}, {Currie}, {Danielson}, {Ewald}, {Faber},
  {Grillmair}, {Groth}, {Holtzman}, {Hunter}, {Kristian}, {Light}, {Lynds},
  {Monet}, {O'Neil}, {Shaya}, {Seidelmann}, \&
  {Westphal}}]{1996AJ....111.2349H}
{Hester}, J.~J., {Scowen}, P.~A., {Sankrit}, R., {Lauer}, T.~R., {Ajhar},
  E.~A., {Baum}, W.~A., {Code}, A., {Currie}, D.~G., {Danielson}, G.~E.,
  {Ewald}, S.~P., {Faber}, S.~M., {Grillmair}, C.~J., {Groth}, E.~J.,
  {Holtzman}, J.~A., {Hunter}, D.~A., {Kristian}, J., {Light}, R.~M., {Lynds},
  C.~R., {Monet}, D.~G., {O'Neil}, E.~J., {Shaya}, E.~J., {Seidelmann}, K.~P.,
  \& {Westphal}, J.~A. 1996, \aj, 111, 2349


\bibitem[{{Israel}(1978)}]{1978A&A....70..769I}
{Israel}, F.~P. 1978, \aap, 70, 769


\bibitem[{{Johnstone} \& {Bally}(1999)}]{1999ApJ...510L..49J}
{Johnstone}, D. \& {Bally}, J. 1999, \apjl, 510, L49
 \href{http://adsabs.harvard.edu/cgi-bin/nph-bib_query?bibcode=1999ApJ...510L.%
.49J&db_key=AST}{\adsurllinklabel}

\bibitem[{{Johnstone} {et~al.}(1998){Johnstone}, {Hollenbach}, \&
  {Bally}}]{1998ApJ...499..758J}
{Johnstone}, D., {Hollenbach}, D., \& {Bally}, J. 1998, \apj, 499, 758
 \href{http://adsabs.harvard.edu/cgi-bin/nph-bib_query?bibcode=1998ApJ...499..%
758J&db_key=AST}{\adsurllinklabel}

\bibitem[{{Jones}(1992)}]{1992PhDT........35J}
{Jones}, M.~R. 1992, Ph.D.~Thesis


\bibitem[{{Kahn}(1954)}]{1954BAN....12..187K}
{Kahn}, F.~D. 1954, \bain, 12, 187


\bibitem[{{Kaler}(1967)}]{1967ApJ...148..925K}
{Kaler}, J.~B. 1967, \apj, 148, 925


\bibitem[{{L{\' o}pez-Mart{\'{\i}}n} {et~al.}(2001){L{\' o}pez-Mart{\'{\i}}n},
  {Raga}, {Mellema}, {Henney}, \& {Cant{\' o}}}]{2001ApJ...548..288L}
{L{\' o}pez-Mart{\'{\i}}n}, L., {Raga}, A.~C., {Mellema}, G., {Henney}, W.~J.,
  \& {Cant{\' o}}, J. 2001, \apj, 548, 288
 \href{http://adsabs.harvard.edu/cgi-bin/nph-bib_query?bibcode=2001ApJ...548..%
288L&db_key=AST}{\adsurllinklabel}

\bibitem[{{Megeath} {et~al.}(1990){Megeath}, {Herter}, {Gull}, \&
  {Houck}}]{1990ApJ...356..534M}
{Megeath}, S.~T., {Herter}, T., {Gull}, G.~E., \& {Houck}, J.~R. 1990, \apj,
  356, 534


\bibitem[{{Moore} {et~al.}(2002){Moore}, {Hester}, {Scowen}, \&
  {Walter}}]{2002AJ....124.3305M}
{Moore}, B.~D., {Hester}, J.~J., {Scowen}, P.~A., \& {Walter}, D.~K. 2002, \aj,
  124, 3305
 \href{http://adsabs.harvard.edu/cgi-bin/nph-bib_query?bibcode=2002AJ....124.3%
305M&db_key=AST}{\adsurllinklabel}

\bibitem[{{Myers} \& {Goodman}(1988)}]{1988ApJ...326L..27M}
{Myers}, P.~C. \& {Goodman}, A.~A. 1988, \apjl, 326, L27


\bibitem[{{O'Dell}(2001{\natexlab{a}})}]{2001PASP..113...29O}
{O'Dell}, C.~R. 2001{\natexlab{a}}, \pasp, 113, 29


\bibitem[{{O'Dell}(2001{\natexlab{b}})}]{2001ARA&A..39...99O}
---. 2001{\natexlab{b}}, \araa, 39, 99
 \href{http://adsabs.harvard.edu/cgi-bin/nph-bib_query?bibcode=2001ARA%26A..39%
...99O&db_key=AST}{\adsurllinklabel}

\bibitem[{{O'Dell} \& {Doi}(1999)}]{1999PASP..111.1316O}
{O'Dell}, C.~R. \& {Doi}, T. 1999, \pasp, 111, 1316
 \href{http://adsabs.harvard.edu/cgi-bin/nph-bib_query?bibcode=1999PASP..111.1%
316O&db_key=AST}{\adsurllinklabel}

\bibitem[{{O'Dell} {et~al.}(2001){O'Dell}, {Ferland}, \&
  {Henney}}]{2001ApJ...556..203O}
{O'Dell}, C.~R., {Ferland}, G.~J., \& {Henney}, W.~J. 2001, \apj, 556, 203
 \href{http://adsabs.harvard.edu/cgi-bin/nph-bib_query?bibcode=2001ApJ...556..%
203O&db_key=AST}{\adsurllinklabel}

\bibitem[{{O'Dell} {et~al.}(1997){O'Dell}, {Hartigan}, {Lane}, {Wong},
  {Burton}, {Raymond}, \& {Axon}}]{1997AJ....114..730O}
{O'Dell}, C.~R., {Hartigan}, P., {Lane}, W.~M., {Wong}, S.~K., {Burton}, M.~G.,
  {Raymond}, J., \& {Axon}, D.~J. 1997, \aj, 114, 730


\bibitem[{{O'Dell} {et~al.}(2000){O'Dell}, {Henney}, \&
  {Burkert}}]{2000AJ....119.2910O}
{O'Dell}, C.~R., {Henney}, W.~J., \& {Burkert}, A. 2000, \aj, 119, 2910
 \href{http://adsabs.harvard.edu/cgi-bin/nph-bib_query?bibcode=2000AJ....119.2%
910O&db_key=AST}{\adsurllinklabel}

\bibitem[{{O'Dell} {et~al.}(2003){O'Dell}, {Peimbert}, \&
  {Peimbert}}]{2003AJ....125.2590O}
{O'Dell}, C.~R., {Peimbert}, M., \& {Peimbert}, A. 2003, \aj, 125, 2590


\bibitem[{{O'Dell} {et~al.}(1993){O'Dell}, {Valk}, {Wen}, \&
  {Meyer}}]{1993ApJ...403..678O}
{O'Dell}, C.~R., {Valk}, J.~H., {Wen}, Z., \& {Meyer}, D.~M. 1993, \apj, 403,
  678


\bibitem[{{O'Dell} {et~al.}(1991){O'Dell}, {Wen}, \&
  {Hester}}]{1991PASP..103..824O}
{O'Dell}, C.~R., {Wen}, Z., \& {Hester}, J.~J. 1991, \pasp, 103, 824


\bibitem[{{O'Dell} \& {Yusef-Zadeh}(2000)}]{2000AJ....120..382O}
{O'Dell}, C.~R. \& {Yusef-Zadeh}, F. 2000, \aj, 120, 382


\bibitem[{{Omodaka} {et~al.}(1994){Omodaka}, {Hayashi}, {Hasegawa}, \&
  {Hayashi}}]{1994ApJ...430..256O}
{Omodaka}, T., {Hayashi}, M., {Hasegawa}, T., \& {Hayashi}, S.~S. 1994, \apj,
  430, 256


\bibitem[{{Osterbrock}(1989)}]{Osterbrock-book}
{Osterbrock}, D.~E. 1989, {Astrophysics of gaseous nebulae and active galactic
  nuclei} (Mill Valley, CA: University Science Books)


\bibitem[{{Pankonin} {et~al.}(1979){Pankonin}, {Walmsley}, \&
  {Harwit}}]{1979A&A....75...34P}
{Pankonin}, V., {Walmsley}, C.~M., \& {Harwit}, M. 1979, \aap, 75, 34


\bibitem[{{Pogge} {et~al.}(1992){Pogge}, {Owen}, \&
  {Atwood}}]{1992ApJ...399..147P}
{Pogge}, R.~W., {Owen}, J.~M., \& {Atwood}, B. 1992, \apj, 399, 147


\bibitem[{{Pottasch}(1956)}]{1956BAN....13...77P}
{Pottasch}, S.~R. 1956, \bain, 13, 77


\bibitem[{Quirk(1994)}]{Quirk1994}
Quirk, J.~J. 1994, Internat. J. Numer. Methods Fluids, 18, 555


\bibitem[{{Raga} {et~al.}(1999){Raga}, {Mellema}, {Arthur}, {Binette},
  {Ferruit}, \& {Steffen}}]{1999RMxAA..35..123R}
{Raga}, A.~C., {Mellema}, G., {Arthur}, S.~J., {Binette}, L., {Ferruit}, P., \&
  {Steffen}, W. 1999, Revista Mexicana de Astronomia y Astrofisica, 35, 123


\bibitem[{{Rao} {et~al.}(1998){Rao}, {Crutcher}, {Plambeck}, \&
  {Wright}}]{1998ApJ...502L..75R}
{Rao}, R., {Crutcher}, R.~M., {Plambeck}, R.~L., \& {Wright}, M.~C.~H. 1998,
  \apjl, 502, L75+


\bibitem[{{Redman} {et~al.}(1998){Redman}, {Williams}, {Dyson}, {Hartquist}, \&
  {Fernandez}}]{1998A&A...331.1099R}
{Redman}, M.~P., {Williams}, R.~J.~R., {Dyson}, J.~E., {Hartquist}, T.~W., \&
  {Fernandez}, B.~R. 1998, \aap, 331, 1099
 \href{http://adsabs.harvard.edu/cgi-bin/nph-bib_query?bibcode=1998A%26A...331%
.1099R&db_key=AST}{\adsurllinklabel}

\bibitem[{{Rubin} {et~al.}(1991){Rubin}, {Simpson}, {Haas}, \&
  {Erickson}}]{1991ApJ...374..564R}
{Rubin}, R.~H., {Simpson}, J.~P., {Haas}, M.~R., \& {Erickson}, E.~F. 1991,
  \apj, 374, 564


\bibitem[{Sanders {et~al.}(1998)Sanders, Morano, \& Druguet}]{Sanders1998}
Sanders, R., Morano, E., \& Druguet, M. 1998, J.~Comp.~Phys., 145, 511


\bibitem[{{Sankrit} \& {Hester}(2000)}]{2000ApJ...535..847S}
{Sankrit}, R. \& {Hester}, J.~J. 2000, \apj, 535, 847
 \href{http://adsabs.harvard.edu/cgi-bin/nph-bib_query?bibcode=2000ApJ...535..%
847S&db_key=AST}{\adsurllinklabel}

\bibitem[{{Scowen} {et~al.}(1998){Scowen}, {Hester}, {Sankrit}, {Gallagher},
  {Ballester}, {Burrows}, {Clarke}, {Crisp}, {Evans}, {Griffiths}, {Hoessel},
  {Holtzman}, {Krist}, {Mould}, {Stapelfeldt}, {Trauger}, {Watson}, \&
  {Westphal}}]{1998AJ....116..163S}
{Scowen}, P.~A., {Hester}, J.~J., {Sankrit}, R., {Gallagher}, J.~S.,
  {Ballester}, G.~E., {Burrows}, C.~J., {Clarke}, J.~T., {Crisp}, D., {Evans},
  R.~W., {Griffiths}, R.~E., {Hoessel}, J.~G., {Holtzman}, J.~A., {Krist}, J.,
  {Mould}, J.~R., {Stapelfeldt}, K.~R., {Trauger}, J.~T., {Watson}, A.~M., \&
  {Westphal}, J.~A. 1998, \aj, 116, 163
 \href{http://adsabs.harvard.edu/cgi-bin/nph-bib_query?bibcode=1998AJ....116..%
163S&db_key=AST}{\adsurllinklabel}

\bibitem[{{Shu}(1992)}]{1992phas.book.....S}
{Shu}, F.~H. 1992, {Physics of Astrophysics, Vol. II} (University Science
  Books)


\bibitem[{{Subrahmanyan} {et~al.}(2001){Subrahmanyan}, {Goss}, \&
  {Malin}}]{2001AJ....121..399S}
{Subrahmanyan}, R., {Goss}, W.~M., \& {Malin}, D.~F. 2001, \aj, 121, 399
 \href{http://adsabs.harvard.edu/cgi-bin/nph-bib_query?bibcode=2001AJ....121..%
399S&db_key=AST}{\adsurllinklabel}

\bibitem[{{Tenorio-Tagle}(1979)}]{1979A&A....71...59T}
{Tenorio-Tagle}, G. 1979, \aap, 71, 59


\bibitem[{{Tenorio-Tagle} \& {Bedijn}(1981)}]{1981A&A....99..305T}
{Tenorio-Tagle}, G. \& {Bedijn}, P.~J. 1981, \aap, 99, 305


\bibitem[{{van der Werf} \& {Goss}(1989)}]{1989A&A...224..209V}
{van der Werf}, P.~P. \& {Goss}, W.~M. 1989, \aap, 224, 209


\bibitem[{van Leer(1982)}]{vanLeer1982}
van Leer, B. 1982, in Lecture Notes in Physics, Vol. 170, xxx, ed. E.~Krause
  (Springer-Verlag), 507


\bibitem[{{Vandervoort}(1963)}]{1963AJ.....68R.296V}
{Vandervoort}, P.~O. 1963, \aj, 68, 296


\bibitem[{{Vandervoort}(1964)}]{1964ApJ...139..869V}
---. 1964, \apj, 139, 869


\bibitem[{{Wen} \& {O'Dell}(1995)}]{1995ApJ...438..784W}
{Wen}, Z. \& {O'Dell}, C.~R. 1995, \apj, 438, 784


\bibitem[{{Williams}(1999)}]{1999MNRAS.310..789W}
{Williams}, R.~J.~R. 1999, \mnras, 310, 789
 \href{http://adsabs.harvard.edu/cgi-bin/nph-bib_query?bibcode=1999MNRAS.310..%
789W&db_key=AST}{\adsurllinklabel}

\bibitem[{{Williams}(2002)}]{2002MNRAS.331..693W}
---. 2002, \mnras, 331, 693
 \href{http://adsabs.harvard.edu/cgi-bin/nph-bib_query?bibcode=2002MNRAS.331..%
693W&db_key=AST}{\adsurllinklabel}

\bibitem[{{Williams} \& {Dyson}(2001)}]{2001MNRAS.325..293W}
{Williams}, R.~J.~R. \& {Dyson}, J.~E. 2001, \mnras, 325, 293
 \href{http://adsabs.harvard.edu/cgi-bin/nph-bib_query?bibcode=2001MNRAS.325..%
293W&db_key=AST}{\adsurllinklabel}

\bibitem[{{Williams} {et~al.}(2000){Williams}, {Dyson}, \&
  {Hartquist}}]{2000MNRAS.314..315W}
{Williams}, R.~J.~R., {Dyson}, J.~E., \& {Hartquist}, T.~W. 2000, \mnras, 314,
  315
 \href{http://adsabs.harvard.edu/cgi-bin/nph-bib_query?bibcode=2000MNRAS.314..%
315W&db_key=AST}{\adsurllinklabel}

\bibitem[{{Wilson} {et~al.}(2001){Wilson}, {Muders}, {Kramer}, \&
  {Henkel}}]{2001ApJ...557..240W}
{Wilson}, T.~L., {Muders}, D., {Kramer}, C., \& {Henkel}, C. 2001, \apj, 557,
  240


\bibitem[{{Yorke} {et~al.}(1982){Yorke}, {Bodenheimer}, \&
  {Tenorio-Tagle}}]{1982A&A...108...25Y}
{Yorke}, H.~W., {Bodenheimer}, P., \& {Tenorio-Tagle}, G. 1982, \aap, 108, 25


\bibitem[{{Zuckerman}(1973)}]{1973ApJ...183..863Z}
{Zuckerman}, B. 1973, \apj, 183, 863


\end{thebibliography}



\appendix

\ifpreprint
  \setkeys{Gin}{width=0.6\linewidth,height=0.5\textheight,keepaspectratio=true}
\fi

\section{Analytic Model for the Flow from a Flat Ionization Front}
\label{sec:analytic-model-flow}

In this section, we develop an approximate analytic treatment of the
ionized portion of the photoevaporation flow, for the special case in
which the ionization front is exactly flat.  An earlier version of
this calculation was presented in \citet{2003RMxAC..15..175H}.

\subsection{Assumptions and Simplifications}
\label{sec:assumpt-simpl}

\begin{figure}\centering
  \includegraphics{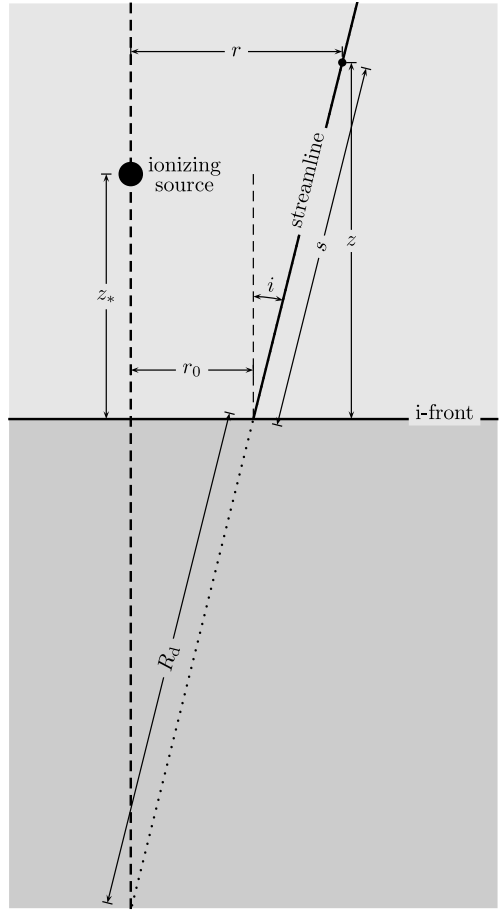}
  \caption{Schematic diagram of the photoevaporation flow from a
    plane ionization front illuminated by a point source, in which
    the principal geometrical quantities are indicated.}
  \label{fig:plane-flow}
\end{figure}

Consider an infinite plane ionization front ($z=0$) illuminated by a
point source of ionizing radiation, which lies at a height $z=z_*$
above the \ifront{} and defines the cylindrical axis $r=0$ (see
Figure~\ref{fig:plane-flow}).
Suppose that the \ifront{} is
everywhere \hbox{D-critical} and that the volume $z>0$ is filled with a
time-steady isothermal ionized
photoevaporation flow with straight streamlines.  Each streamline can
be labeled with the cylindrical radius $r_0$ of its footpoint on the
\ifront{} and makes an angle $i(r_0)$ with the $z$-direction. Assume
that the acceleration of the flow along a streamline  is identical to
that in the photoevaporation flow from a spherical globule with a
radius equal to the local ``divergence radius'' of the flow: $R\Sub{d}
= r_0 \csc i$ so that the velocity along each streamline follows an
identical function $U(y) \equiv u/c\Sub{i}$, where $y = 1 +
s/R\Sub{d}$, $s$ is the distance along the streamline, and $c\Sub{i}$
is the isothermal sound speed in the ionized gas (since the front is
\hbox{D-critical}, we have $U(0) = 1$).  The density at any point ($r$, $z$)
can therefore be separated as
\begin{equation}
  \label{eq:density-decomposition}
  n(r,z) = \frac{ N(r_0) } {y^2 U(y)}
\end{equation}
where $r = r_0 + (y-1) R\Sub{d} \sin i$, $z = (y-1) R\Sub{d} \cos i$,
and $N(r_0)$ is the density at the ionization front.

The effective recombination thickness, $h$, is defined as in the
caption to Figure~\ref{fig:lateral}, except integrated along a
streamline rather than along the $z$-axis: $ N^2(r_0) h =
\int_0^\infty n^2 (r_0,s) \, ds$, yielding
\begin{equation}
  \label{eq:h-definition}
  h = R\Sub{d} \int_1^\infty y^{-4} U^{-2} \, dy = \omega R\Sub{d} , 
\end{equation}
where $\omega$ is a constant that depends only on the acceleration law
$U(y)$. Although $\omega \simeq 0.1$ is typical for spherical
photoevaporation flows from globules \citep{1989ApJ...346..735B}, the
results of our numerical models above (see Fig.~\ref{fig:evo-kappa})
suggest that the acceleration is somewhat slower in the present
circumstances, leading to a larger value of $h$, so we leave $\omega$
as a free parameter for the time being.

\subsection{Photoionization Balance}
\label{sec:phot-balance}

With this definition of $h$, and assuming that the photoevaporation
flow is ``recombination dominated'' \citep{2001RMxAC..10...57H}, that
recombinations can be treated in the on-the-spot approximation, and
ignoring any absorption of ionizing photons by grains, one can
approximate the recombination/ionization balance along a line from the
ionizing source to the \ifront{} by considering the perpendicular flux
of ionizing photons incident on the equivalent homogeneous layer:
\begin{equation}
  \label{eq:ionization-balance}
  \frac{ Q\Sub{H} \cos \theta }{ 4\pi (z_*^2 +r_0^2) } = 
  \alpha\Sub{B} N^2(r_0) h (r_0) , 
\end{equation}
where $\tan\theta = r_0 / z_*$, $Q\Sub{H}$ is the ionizing photon
luminosity (s$^{-1}$) of the source, and $\alpha\Sub{B}$ is the Case~B
recombination coefficient. This approximation will be valid when $h$
is small enough to satisfy $h < z_*$ and $h\sec \theta < H$, where $H
\equiv \left|d \ln N / d r_0\right|^{-1}$ is the radial scale length
of the ionized density profile at the \ifront{}. This profile can be
found from equation~(\ref{eq:ionization-balance}) to be
\begin{equation}
  \label{eq:density-profile}
  N(r_0) = N_0 \left(1+\frac{r_0^2}{z_*^2}\right)^{-3/4} \left(\frac{h}{h_0}\right)^{-1/2}, 
\end{equation}
where $x = r_0 / z_*$ and $N_0$, $h_0$ are the values on the axis at
$r_0 = 0$. Note that for any plausible $h(r_0)$, the density at the
\ifront{} will decrease with increasing radius.

\subsection{Lateral Acceleration}
\label{sec:lateral-acceleration}

In order to complete the solution, it is now only necessary to specify
the radial dependence of the streamline angle, $i(r_0)$, from which
the radius of divergence, $R\Sub{d}(r_0)$ and scale height, $h(r_0)$,
follow automatically. To do this, we must recognise that the
streamlines cannot truly be straight: in a \hbox{D-critical} front the gas
will be accelerated up to the ionized sound speed, $c\Sub{i}$, in the
$z$-direction (perpendicular to the front) but at $z=0$ it will have
no velocity in the $r$-direction (parallel to the front). However, the
pressure gradient associated with the radially decreasing density will
laterally accelerate the gas in the positive $r$-direction as it rises
through the recombination layer (which is much thicker than the front
itself), thus bending the streamlines outwards. To be definite, we
will consider the characteristic angle of each streamline, $i(r_0)$,
to be that obtained by the gas at $z = h(r_0)$.

The lateral acceleration, $a$, of the gas is given by 
\begin{equation}
  \label{eq:acceleration}
  a = \frac{1}{\rho} \frac{d P}{d r} = \frac{ c\Sub{i}^2 }{ H } , 
\end{equation}
where $H$ is the lateral density scale length defined above. 
Immediately after passing through the heating front, the gas will have
a vertical velocity of $\simeq 0.3 c_\mathrm{i}$, rising to $\simeq
c_\mathrm{i}$ once it is fully ionized and then to $\simeq 2
c_\mathrm{i}$ as it passes through the recombination layer. We take
the typical vertical speed to be $v_z = \beta c_\mathrm{i}$, where $\beta$
is a parameter whose value must be determined but which should be
somewhat less than unity. Hence,  the time taken to rise
to a height $h$ is $\delta t = h/\beta c\Sub{i}$, so that the lateral speed reached is
$v_r = a \delta t = c\Sub{i} h / \beta H$ and the streamline angle is given by 
\begin{equation}
  \label{eq:streamline-angle}
  \tan i = \frac{v_r}{v_z} = \frac{h}{\beta H} ,
\end{equation}
so that $R\Sub{d} = h / \omega = r_0 ( 1 + \beta^2 H^2/h^2 )^{1/2}$
(see Fig.~\ref{fig:plane-flow}). In the
limit that $ h^2/\beta^2 H^2 \ll 1$, this can be solved to give
\begin{equation}
  \label{eq:h-in-H-r0}
  h \simeq (\omega \beta H r_0)^{1/2} ,
\end{equation}
which can be combined with equation~(\ref{eq:density-profile}) and the
definition of $H$ to give the following ODE for $h$:
\begin{equation}
  \label{eq:h-ODE}
  \frac{dh}{dr_0} + \frac{3hr_0}{r_0^2+z_*^2} - \frac{2\omega\beta r_0}{h}
  = 0 .
\end{equation}
This can be solved directly by means of the substitution $y = h^2$ and
the integrating factor $(r_0^2+z_*^2)^3$ to give
\begin{equation}
  \label{eq:h-final}
  h = \left(\frac{\omega\beta}2\right)^{1/2} z_0 \left( 1 + \frac{r_0^2}{z_*^2} \right)^{1/2}
\end{equation}
which may in turn be substituted back into
equation~(\ref{eq:density-profile}) to give
\begin{equation}
  \label{eq:density-final}
  N(r_0) = N_0 \left(1+\frac{r_0^2}{z_*^2}\right)^{-1} . 
\end{equation}
From equation~(\ref{eq:streamline-angle}), the streamline angles are
then given by 
\begin{equation}
  \label{eq:streamline-final}
  \tan i =  \left(\frac{\omega}{2\beta}\right)^{1/2}
  \frac{r_0}{\left(r_0^2+z_*^2\right)^{1/2}} .   
\end{equation}

Hence, the vertical scale height on the axis is given by $h_0 =
(\omega\beta/2)^{1/2} z_*$, at which point the streamlines are
vertical. We can now fix the value of $\omega\beta$ by comparing this
with our numerical results from the model simulations.
Figure~\ref{fig:evo-kappa} shows that for zero curvature,
$h_\mathrm{eff}/z_\mathrm{h} \simeq 0.4$, which implies $\omega\beta
\simeq 0.3$.  Moving away from the axis, the streamlines start to
diverge slightly, reaching an asympotic angle of $ i_\infty =
\tan^{-1} (\omega/2\beta)^{1/2}$ as $r_0 \rightarrow \infty$. In
Figure~\ref{fig:streamcomp} this behavior of the streamline angle is
compared with that seen in our numerical simulations of Model~B, where
it is found that $\omega/\beta \simeq 0.6$ gives a reasonable fit for
$r_0 < z_*$, which, combined with the estimate of $\omega\beta$,
yields $\omega \simeq 0.4$ and $\beta \simeq 0.7$.

\begin{figure}
  \centering
  \includegraphics{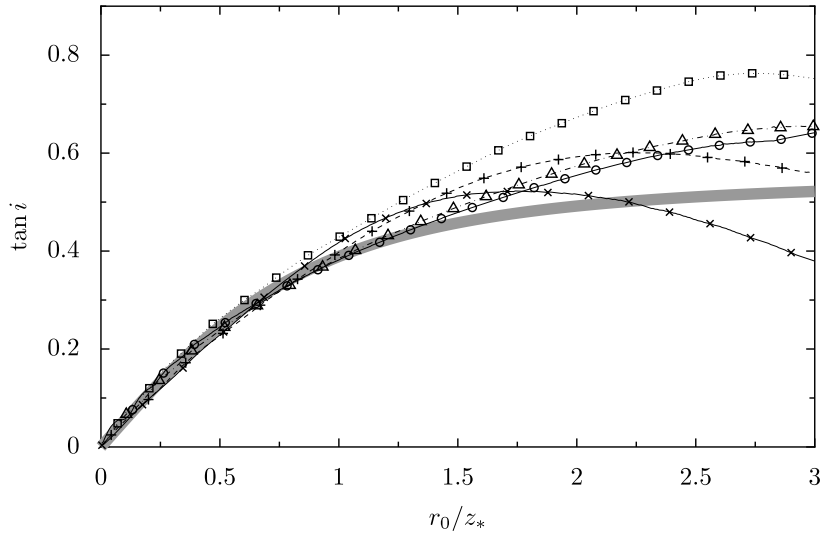}
  \caption{Comparison of the increase in streamline angle with radius
    between the analytic model (thick gray line) and the numerical
    simulation of Model~B at various evolutionary times: 40,000
    (crosses), 60,000 (plusses), 80,000 (squares), 100,000 (circles),
    and 120,000 (triangles) years.}
  \label{fig:streamcomp}
\end{figure}

The validity of the analytical results for $r_0$ greater than a few
times $z_*$ is not to be expected, since $h$ no longer satisfies the
conditions stated after equation~(\ref{eq:ionization-balance}) above.
Nonetheless, it is satisfying that the general characteristics of the
photoevaporation flows seen in our numerical simulations can be
reproduced from these simple physical arguments.

It is interesting to consider how the calculation would be affected by
the presence of dust grains in the ionized flow, which absorb part of
the ionizing radiation. To first order, this may be treated by
considering the column density of gas between the ionizing star and
the ionization front, which is proportional to $N h / \cos \theta$.
From equations~(\ref{eq:h-final}) and~(\ref{eq:density-final}) it can
be seen that this column density is a constant for all values of
$r_0$. As a result, dust absorption has no \emph{differential} effect
on the structure of the flow in this approximation but merely reduces
the effective ionizing luminosity of the star.

\section{Are the Ionization Fronts D-critical?}
\label{sec:class-ioniz-front}

\begin{figure}\centering
  \includegraphics{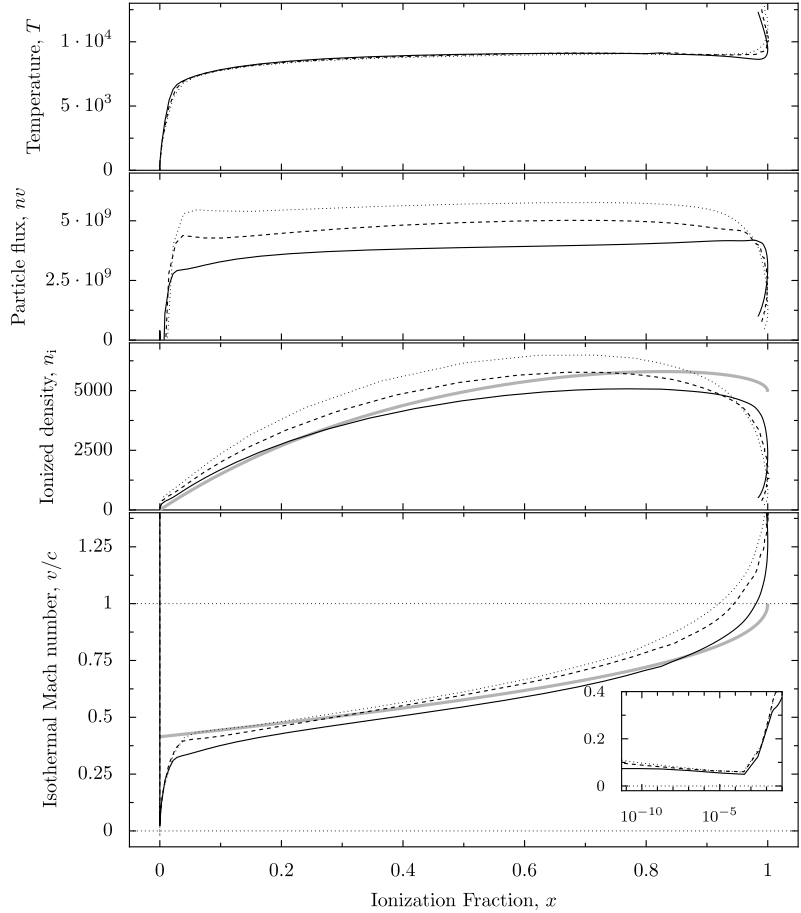}
\caption{Structure of the ionization front: various physical
    quantities as a function of ionization fraction, $x$.
    (\textit{a})~Gas temperature. (\textit{b})~Particle flux.
    (\textit{c})~Ionized density.  (\textit{d})~Isothermal Mach
    number, with inset showing the same quantity on a logarithmic
    scale for small values of $x$. The different models are shown by:
    A (solid line), B (dashed line), and C (dotted line). The analytic
    curves of ionized density and Mach number for a steady,
    plane-parallel, isothermal, \hbox{D-critical} front are shown by thick
    gray lines. All velocities are given relative to the heating
    front.}
  \label{fig:all-vs-ifrac}
\end{figure}

The traditional classification of ionization fronts into
weak/critical/strong and R-type/D-type
\citep{1954BAN....12..187K,Goldsworthy-1961} was carried out for
plane-parallel fronts. Although the fronts can be curved in our
simulations, the ionization front thickness is always much less than
the radius of curvature and so a plane-parallel approximation is
justified for the front itself. It is therefore an interesting
question whether or not our fronts are indeed \hbox{D-critical}, as is often
assumed for photoevaporation flows. 

Two approaches are possible to the empirical classification of the
ionization fronts in the simulations. One possibility is to look at
the velocity of propagation of the front with respect to the upstream
gas.  The other is to look at the internal structure of the front,
which is the method we investigate first.

Figure~\ref{fig:all-vs-ifrac} graphs various quantities as a function
of ionization fraction, $x$. Figure~\ref{fig:all-vs-ifrac}\textit{a}
shows that almost all the gas heating in the front occurs at very low
values of $x$, so the ionization transition itself can be considered
approximately isothermal. Furthermore,
Figure~\ref{fig:all-vs-ifrac}\textit{b} shows that the particle flux
is roughly constant through a great part of the front. At very low
ionization fractions, the flow is not steady because the shock
propagates at a slightly faster speed than the ionization front. At
high ionization fractions, the streamlines start to diverge and the
flow is no longer plane parallel, resulting in a reduction of the
particle flux. Nonetheless, the flow can be seen to be approximately
steady and plane-parallel over virtually the entire front. 

In this case, one can apply the simple model developed in Appendix~A
of \citet{2005ApJ...621..328H}. For a time-steady, plane-parallel,
isothermal, \hbox{D-critical} front, where the critical point occurs at
$x=1$, then the results of that paper imply
\begin{equation}
  \label{eq:mach-dcrit}
  M(x) = \frac{ 2^{1/2} - (1-x)^{1/2} }{ (1+x)^{1/2} }
\end{equation}
and
\begin{equation}
  \label{eq:di-dcrit}
  n_\mathrm{i} \propto \frac{x}{1-[(1-x)/2]^{1/2}} .
\end{equation}
These curves are shown in Figure~\ref{fig:all-vs-ifrac}\textit{c} and
\textit{d} (thick gray lines), together with the same quantities for
the three numerical models at an age of $78,000\,\years$. The
normalization of the theoretical $n_\mathrm{i}$ curve is arbitrary.
The three models are very similar and also agree quite well with the
analytical curve, except for near $x=0$ and $x=1$. In particular, the
\hbox{D-critical} curve implies that the Mach number immediately after the
heating front at $x\simeq0$ should be $M\Sub{h} = \sqrt{2}-1 \simeq
0.414$, which is close to that observed in the numerical models. This
implies that the models are all close to \hbox{D-critical}, although they all
pass through the sonic point before reaching $x = 1$, with the most
divergent model (C) doing so at $x\simeq0.9$. Model~A, which is the
least divergent, shows a somewhat smaller initial jump in the Mach
number, which may be evidence that it is weaker than \hbox{D-critical}.

The numerical models all show a maximum in the ionized density between
$x=0.6$ and $x=0.8$, similar to that predicted by the \hbox{D-critical}
curve. At higher values of $x$, on the other hand, the ionized density
decreases much faster than in the analytical model because of the
divergence in the numerical models. Note, however, that the particle
flux is constant in the vicinity of the maxima in $n_\mathrm{i}$, so
divergence is not important there. 

\begin{figure}\centering
  \includegraphics{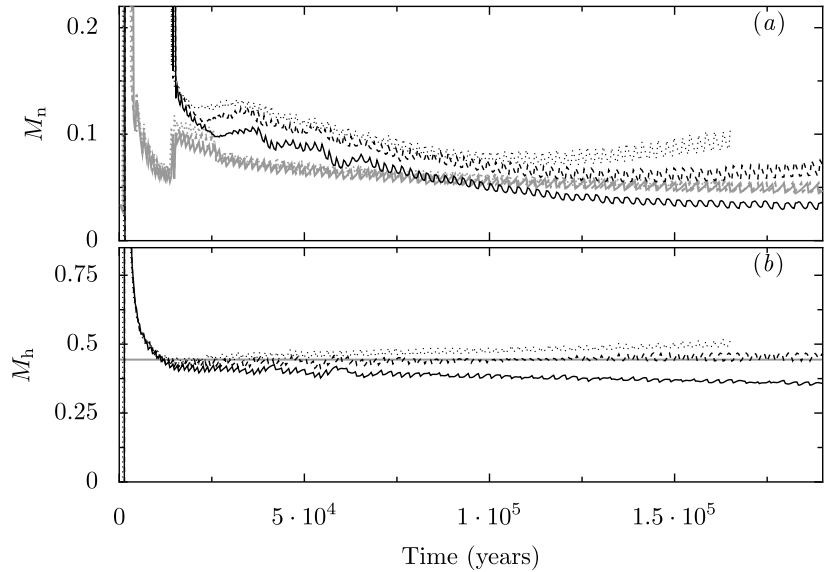}
  \caption{Evolution with time of relative Mach numbers of the gas
    flow either side of the heating front. (\textit{a})~Relative Mach
    number with which neutral gas enters the heating front,
    $M_\mathrm{n}$ (black lines). Also shown (gray lines) is the
    critical Mach number for a steady flow,
    $c_\mathrm{n}/2c_\mathrm{i}$.  (\textit{b})~Relative Mach number
    with which gas leaves the heating front, $M_\mathrm{h}$,
    calculated at $x=0.1$. Gray line shows the analytical
    approximation for a \hbox{D-critical} front.}
  \label{fig:evo-mach}
\end{figure}

The other test of the \hbox{D-critical} hypothesis is to look at the Mach
number on the neutral side, that is, the Mach number of the gas flow
relative to the heating front as it enters the front from the shocked
neutral shell, $M_\mathrm{n}$. For a steady, plane-parallel front,
there exists a maximum value, $M_\mathrm{n,max} =
c_\mathrm{n}/2c_\mathrm{i}$, which corresponds to a \hbox{D-critical} front.
$M_\mathrm{n}$ is about 0.06 for the models shown in
Fig.~\ref{fig:all-vs-ifrac}\textit{d}, as can be seen from the inset.
The evolution with time of $M_\mathrm{n}$ for the models is shown in
Figure~\ref{fig:evo-mach}\textit{a}, which also shows (gray line) the
theoretical maximum value, $M_\mathrm{n,max}$, taking $c_\mathrm{i}$
to be the sound speed at the sonic point in the flow. It can be seen
that $M_\mathrm{n}$ actually exceeds $M_\mathrm{n,max}$ everywhere
except for the late-time evolution of Model~A! This should not be
possible for a steady flow but is probably due to the fact that the
sound speed in the neutral shell, $c_\mathrm{n}$, varies spatially
within the shell, according to the past history of the propagation
speed of the shock front, $U_\mathrm{s}$, thus violating the
steady-state assumption even though the flow from the shell through
the ionization front \emph{is} approximately steady.  Hence,
$M_\mathrm{n}$ is not a suitable diagnostic for classifying the
ionization front.

The evolution with time of $M_\mathrm{h}$ is shown in
Figure~\ref{fig:evo-mach}\textit{b}, calculated at the point where the
gas leaves the heating front at $x=0.1$. At this point in the flow,
the gas has already heated up to $\simeq 10^4 \,\Kelvin$, although it
is still largely neutral, see Figure~\ref{fig:all-vs-ifrac}. The
analytic value for a \hbox{D-critical} front with a sonic point at $x=1$ is
shown by the gray line (see eq.(\ref{eq:mach-dcrit}) above). The
numerical results for Model~B can be seen to trace the analytical line
almost exactly, indicating that the front is indeed \hbox{D-critical} in this
model.  For Model~C, $M_\mathrm{h}$ slightly exceeds that analytical
value, but this can be explained by the fact that the sonic point
occurs nearer to $x=0.9$ in this model (see
Fig.~\ref{fig:all-vs-ifrac}\textit{d}). For Model~A, $M_\mathrm{h}$
falls increasingly below the \hbox{D-critical} value as the evolution
progresses, indicating that the ionization front becomes increasingly
weak-D.  This implies that there should be no sonic point within the
ionization front itself. This is consistent with the fact that the
sonic point occurs much further from the heating front than in the
other two models (see Fig.~\ref{fig:lateral}\textit{c}), in a region
where the gas is fully ionized.

\end{document}